\documentclass[useAMS,usenatbib]{mn2e}
\usepackage{graphicx}

\newcommand{\Pspin}{\mbox{$P_{\mathrm{spin}}$}}
\newcommand{\Porb}{\mbox{$P_{\mathrm{orb}}$}}
\newcommand{\Line}[3]{\Ion{#1}{#2}\,$\lambda$\,#3}
\newcommand{\Lines}[3]{\Ion{#1}{#2}\,$\lambda\lambda$\,#3}
\newcommand{\Ion}[2]{#1{\,\scriptsize #2}}
\newcommand{\id}{\mbox{$\mathrm{d^{-1}}$}}
\newcommand{\kms}{\mbox{$\mathrm{km\,s^{-1}}$}}
\newcommand{\cts}{\mbox{$\mathrm{cts\;s^{-1}}$}}
\newcommand{\Nh}{\mbox{$N_{\mathrm{H}}$}}

\title[Four new intermediate polars]{Cataclysmic variables from a
  ROSAT/2MASS selection I: Four new intermediate polars}

\author[B.T. G\"ansicke et al.]{
B. T. G\"ansicke$^1$,
T.R. Marsh$^1$,
A. Edge$^2$,
P. Rodr\'{\i}guez-Gil$^{1,3}$,
D. Steeghs$^4$, \newauthor
\ S. Araujo-Betancor$^5$, 
E. Harlaftis$^6$,
O. Giannakis$^7$,
S. Pyrzas$^8$,
L. Morales-Rueda$^9$,\newauthor
A. Aungwerojwit$^1$\\
$^{1}$ Department of Physics, University of Warwick, Coventry CV4 7AL,
UK \\
$^{2}$ Department of Physics, University of Durham, South Road, Durham DH1
3LE, UK\\
$^{3}$ Instituto de Astrof\'isica de Canarias, 38200 La Laguna, Tenerife, Spain\\
$^{4}$ Harvard-Smithsonian Center for Astrophysics, 60 Garden Street,
MS-67, Cambridge, MA 02138, USA\\
$^{5}$ Space Telescope Science Institute, 3700 San Martin Drive,
Baltimore, MD21218, USA\\
$^{6}$ Institute of Space Applications and Remote Sensing,
   National Observatory of Athens, P.O. Box 20048, Athens 11810,
   Greece \\
$^{7}$ Institute of Astronomy,
   National Observatory of Athens, P.O. Box 20048, Athens 11810,
   Greece \\
$^{8}$ Department of Physics, Section of Astrophysics, Astronomy \&
   Mechanics, University of Thessaloniki, 541 24 Thessaloniki, Greece \\
$^{9}$  Department of Astrophysics, Radboud University Nijmegen, PO Box 9010,
6500GL, Nijmegen, The Netherlands\\
}

\begin{document}

\date{Accepted 2005. Received 2005; in original form 2005}

\pagerange{\pageref{firstpage}--\pageref{lastpage}} \pubyear{2005}

\maketitle

\label{firstpage}

\begin{abstract}
We report the first results from a new search for cataclysmic
variables (CVs) using a combined X-ray (ROSAT) / infrared (2MASS)
target selection that discriminates against background
AGN. Identification spectra were obtained at the Isaac Newton
Telescope for a total of 174 targets, leading to the discovery of 12
new CVs. Initially devised to find short-period low-mass-transfer CVs,
this selection scheme has been very successful in identifying new
intermediate polars. Photometric and spectroscopic follow-up
observations identify four of the new CVs as intermediate polars:
1RXS\,J063631.9+353537 ($\Porb\simeq201$\,min, $\Pspin=1008.3408$\,s or
930.5829\,s), 1RXS\,J070407.9+262501 ($\Porb\simeq250$\,min,
$\Pspin=480.708$\,s), 1RXS\,J173021.5--055933 ($\Porb=925.27$\,min,
$\Pspin=128.0$\,s), and 1RXS\,J180340.0+401214 ($\Porb=160.21$\,min,
$\Pspin=1520.51$\,s). RX\,J1730, also a moderately bright hard X-ray
source in the INTEGRAL/IBIS Galactic plane survey, resembles  the
enigmatic AE\,Aqr. It is likely that its white dwarf is not rotating
at the spin equilibrium period, and the system may represent a
short-lived phase in CV evolution. 
\end{abstract}

\begin{keywords}
accretion, accretion discs -- binaries: close -- novae, cataclysmic variables
\end{keywords}

\section{Introduction}
The evolution of cataclysmic variable stars (CVs) is driven by the
need for the mass-donor stars to fill their Roche lobes.  This is
ensured by angular momentum loss (AML), through gravitational wave
radiation and magnetic braking via a stellar wind, which gradually
decreases the binary orbital period \Porb. The standard theory of CV
evolution assumes that magnetic braking ceases once the systems reach
$\Porb\simeq3$\,h \citep[e.g.][]{rappaportetal83-1}. With
gravitational wave radiation being a much less efficient AML mechanism
than magnetic braking, CVs with $\Porb<3$\,h have much lower mass
transfer rates and longer evolution time scales than those with
$\Porb>3$\,h \citep{kolb+stehle96-1}.  Initially the mass donor
shrinks in response to mass loss causing the orbital period to
decrease. Eventually the mass donor becomes so low in mass ($<
0.08\,\,{\rm M}_\odot$) that fusion dies out and the star evolves
towards a degenerate state. Its radius then starts to increase upon
mass loss, leading to an increase in the orbital period of the CV
\citep{paczynski+sienkiewicz83-1}. Up to this point the decreasing
period is compensated by the reduced mass of the donor bringing about
a near-constant mass loss rate. After the period minimum
(observationally found near 80\,min) however, the simultaneous
reduction in the mass of the donor and the increase in orbital period
act to reduce AML resulting in a precipitous drop in the mass loss
rate, and a further slowing of the CV's evolution. Consequently, the
vast majority ($\sim99$\%) should have periods $<2$\,h, and a large
fraction of all CVs ($\sim 70$\%) should have passed through the
period minimum \citep{kolb93-1, howelletal97-1}.

The apparent lack of observed short-period CVs in general and
``post-period-bounce'' CVs in particlar is a major embarrassment for
our understanding of compact binary evolution
\citep[e.g.][]{gaensickeetal02-2}. Selection effects are an obvious
and generally welcomed explanation for this discrepancy~--~about half
of all known CVs have been found because of strong variability,
i.e. displaying outbursts \citep{gaensicke05-1}. As the absolute
magnitudes of CVs during outburst are a function of their orbital
period, but \textit{not} their mean mass transfer rate
\citep{warner87-1}, there is no obvious reason why post-period-bounce
systems should not be found unless they very rarely or never
go into outburst. Observationally, a number of systems are known that
have very infrequent outbursts (e.g. WZ\,Sge, GW\,Lib, or LL\,And), or
have had no observed outbursts (e.g. BW\,Scl, GD\,552, HS\,2331+3905, or GP\,Com). In
fact, disc instability theory predicts that outburst activity may
cease completely below a certain mass transfer rate
\citep{meyer-hofmeisteretal98-2}. Thus, if the predicted large number
of short-period low-mass-transfer CVs exists, they must be discovered
by other means than variability. Spectroscopic surveys have
demonstrated the ability to find such objects [the Hamburg Quasar
Survey \citep{gaensickeetal02-2}, the Sloan
Digital Sky Survey (\citealt{szkodyetal05-1} and references therein),
and the 2dF Quasar Survey (\citealt{marshetal02-1})] , but the number
of new short orbital period CVs discovered in this way also falls
short of the predictions. We have therefore initiated a new
search for CVs based on one apparently common feature of
low-mass-transfer systems: X-ray emission~--~of the six systems named
above, WZ\,Sge, BW\,Scl, GD\,552, and GP\,Com are noticeable X-ray
sources \citep{vogesetal99-1}, indeed, BW\,Scl was discovered in
X-rays \citep[\,=\,RX\,J2353.0-3852,][]{abbottetal97-1}. Here we
report the first results from this search~--~the discovery of four new
intermediate polars (IPs).

\begin{figure}
\includegraphics[angle=-90,width=\columnwidth]{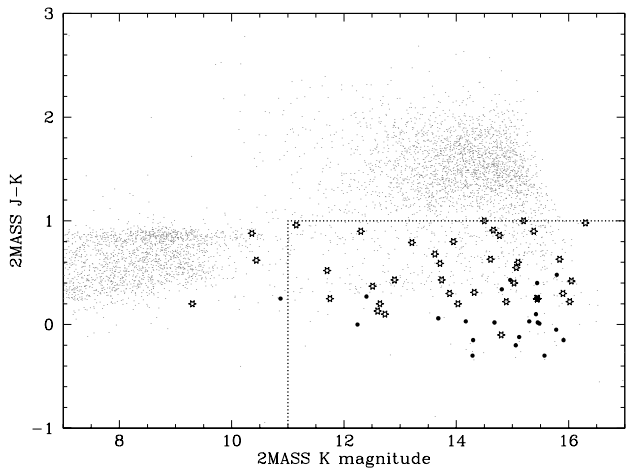}
\caption{\label{f-selection} The 2MASS ``magnitude-colour'' diagram of
stellar (point-like) sources from the ROSAT All Sky Survey with $|b| >
5^\circ$ showing how infrared colours distinguish CVs and white dwarfs
from other sources. Bright stars appear towards the left, while AGN
appear in the upper-right. Known white dwarfs are marked by black dots,
known CVs by stars.  There are 399 objects in the region $K > 11$,
$J-K < 1.0$ designed to avoid the bright stars and AGN. This drops to
300 applying $R < 17$. About 50 of these are known leaving 250
unidentified targets.}
\end{figure}

\section{Target Selection}
\label{s-selection}
X-ray identification work has led to the discovery of various CVs not
found by any other method \citep[e.g.][]{schwopeetal02-1} and the
ROSAT All-Sky Survey (RASS, \citealt{vogesetal99-1}) still contains a
large number of optically unidentified sources, particularly at
intermediate Galactic latitudes ($4^\circ<|b|<30^\circ$). The identification
of CVs is complicated by the fact that they are rare and have similar
optical magnitudes and colours to the more numerous AGN. However, when
near-infrared photometry from 2MASS is also considered there is a
clear discrimination between the bulk of the AGN and CVs or white
dwarfs (Fig.\,\ref{f-selection}). We have drawn the X-ray sources for
this project from the ROSAT Bright Source Catalogue
\citep{vogesetal99-1}, limiting the selection to count rates from 0.05
to 0.2 \cts, a regime not systematically explored to date. There are
over 10\,000 such sources, but applying limitations on galactic
latitude ($|b|>5^\circ$) to avoid confusion, infrared colours ($J-K<1$
consistent with galactic sources) and infrared magnitude ($K>11$ to
exclude the majority of coronal stellar sources) leaves 399 sources.
This drops to 300 with the additional restriction of $R<17$ to allow
idenitifcation work with 2\,m class telescopes. Since some 50 of these
are known objects, mainly CVs or white dwarfs, 250 unidentified
targets remain.

\begin{table}
\begin{minipage}{\columnwidth}
\caption{\label{t-obslog}Log of the observations}
\setlength{\tabcolsep}{1.1ex}
\begin{tabular}{llcccc}
\hline
Date & UT      &  Obs & Filter/Grism & Exp.  & Frames \\
\hline
\multicolumn{2}{l}{\textbf{1RXS\,J063631.9+353537}} \\
2003 Mar 17 & 21:22          & INT & R400V & 900 &  1\\
2004 Oct 21 & 02:42 -- 03:33 & KY & clear & 30  & 195\\
2004 Oct 22 & 00:07 -- 03:43 & KY & clear & 20  & 510\\
2004 Oct 22 & 23:46 -- 03:50 & KY & clear & 20  & 317\\
2004 Oct 23 & 23:12 -- 00:37 & KY & clear & 20  & 187\\
2005 Jan 02 & 00:25 -- 05:42 & WHT & R600B/R316R & 930 & 20 \\
2005 Feb 21 & 20:46 -- 00:52 & KY & clear & 45  & 254\\
2005 Feb 23 & 17:46 -- 01:43 & KY & clear & 45  & 484\\
2005 Feb 24 & 17:16 -- 22:17 & KY & clear & 45  & 352\\
2005 Mar 03 & 17:55 -- 23:53 & KY & clear & 45  & 342\\
\noalign{\smallskip}
\multicolumn{2}{l}{\textbf{1RXS\,J070407.9+26250}} \\
2003 Mar 18 & 21:16          & INT & R400V & 1800 &  1\\
2004 Dec 07 & 02:32 -- 04:43 & KY  & clear & 45  &  144\\
2005 Jan 02 & 00:30 -- 06:49 & INT & clear & 60  &  254\\
2005 Jan 05 & 23:57 -- 05:27 & TNG & clear & 30  &  511\\
2005 Feb 11 & 21:52 -- 02:49 & CA22 & G100 & 600 & 21 \\
\noalign{\smallskip}
\multicolumn{2}{l}{\textbf{1RXS\,J173021.5--055933}} \\
2002 Aug 22 & 22:49          & INT & R400V &  900&  1\\
2003 Apr 10 & 03:00 -- 05:07 & INT & R1200B   & 1500&  6\\
2003 Apr 11 & 03:29 -- 05:16 & INT & R1200B   & 1500&  5\\
2003 Apr 13 & 02:27 -- 05:11 & INT & R1200B   & 1500&  7\\
2003 Apr 22 & 03:29 -- 04:22 & INT & R632V &  600&  6\\
2003 Apr 26 & 03:29 -- 04:22 & INT & R632V &  600&  6\\
2003 Apr 27 & 04:26 -- 05:19 & INT & R632V &  600&  6\\
2003 May 10 & 01:09 -- 05:41 & JKT & $R$   &   30&  339 \\
2003 May 21 & 00:33 -- 04:23 & JKT & $R$   & 15--60&  331 \\
2003 May 20 & 01:49 -- 01:59 & INT & R632V &  600&  2\\
2003 Jun 21 & 23:44 -- 02:40 & CA22 & G100 &  600&  15\\
2003 Jun 22 & 23:21 -- 03:06 & CA22 & G100 &  600&  15\\
2003 Jun 23 & 05:11 -- 07:55 & MAG  & G500/5000 &  120&  27\\
2003 Jun 24 & 03:32 -- 08:15 & MAG  & G500/5000  &  120&  15\\
2003 Jun 25 & 02:39 -- 03:15 & MAG  & G500/5000  &  120&  16\\
2003 Jul 07 & 21:51 -- 03:12 & OGS & clear & 20  &  631\\
2003 Jul 08 & 21:53 -- 03:17 & OGS & clear & 10  & 1030\\
2003 Jul 09 & 21:31 -- 22:31 & OGS & $I$   & 25  &  112\\
2003 Jul 09 & 22:41 -- 23:41 & OGS & $R$   & 25  &  113\\
2003 Jul 09 & 23:52 -- 00:51 & OGS & $V$   & 30  &   95\\
2003 Jul 14 & 21:16 -- 02:09 & OGS & clear & 15  &  777\\
2003 Jul 15 & 21:14 -- 02:01 & OGS & clear & 10  & 1001\\
2003 Jul 16 & 21:30 -- 03:23 & OGS & clear & 12  &  765\\
\noalign{\smallskip}
\multicolumn{2}{l}{\textbf{1RXS\,J180340.0+401214}} \\
2002 Aug 21 & 21:27          & INT & R400V & 1800&  1\\
2002 Aug 28 & 23:26          & INT & R400V & 1800&  1\\
2003 Apr 23 & 04:41 -- 05:34 & INT & R632V & 600 &  6\\
2003 Apr 25 & 04:39 -- 05:42 & INT & R632V & 600 &  7\\
2003 Apr 29 & 02:55 -- 03:59 & CA22& G100  & 600 &  6\\
2003 May 02 & 03:46 -- 04:20 & CA22& G100  & 600 &  6\\
2003 May 19 & 04:15 -- 05:08 & INT & R632V & 600 &  6\\
2003 May 20 & 04:24 -- 05:17 & INT & R632V & 600 &  6\\
2003 May 21 & 22:41 -- 04:13 & JKT & $V$   & 80--100 &  186 \\
2003 Jun 27 & 02:19 -- 03:57 & CA22& G100  & 600 &  9\\
2003 Jun 28 & 01:45 -- 03:57 & CA22& G100  & 600 &  12\\
2003 Aug 15 & 18:26 -- 22:49 & KY  & clear & 30  &  450\\
2003 Aug 16 & 18:30 -- 23:02 & KY  & clear & 30  &  498\\
2003 Aug 17 & 18:13 -- 21:51 & KY  & clear & 30  &  400\\
2003 Aug 18 & 18:08 -- 22:18 & KY  & clear & 30  &  440\\
2003 Aug 19 & 18:17 -- 21:55 & KY  & clear & 30  &  400\\
\hline
\end{tabular}
\end{minipage}
\end{table}

\begin{figure*}
\includegraphics[angle=-90,width=\textwidth]{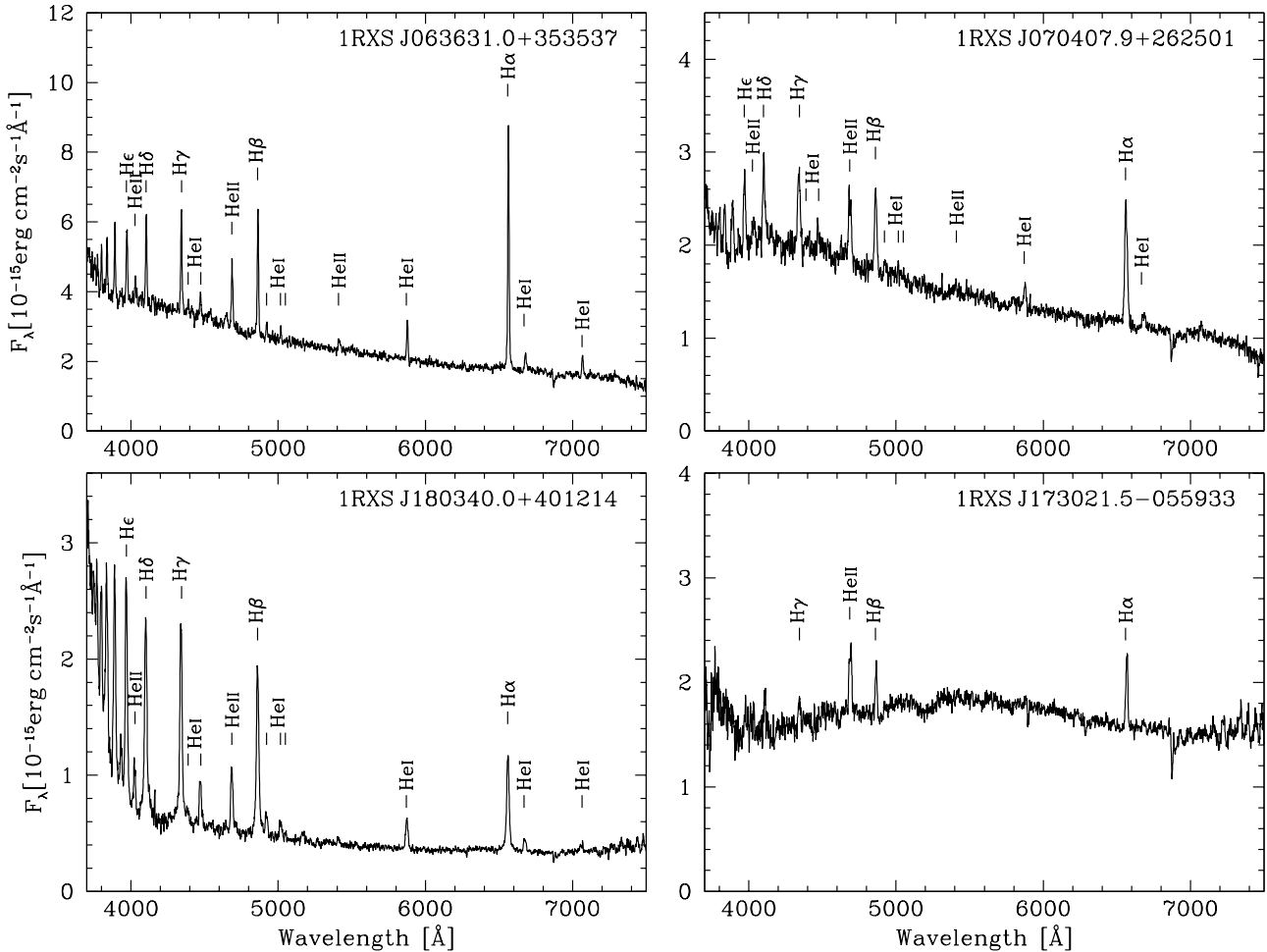}
\caption{\label{f-idspec} INT/IDS identification spectra of RX\,J0636,
RX\,J0704, RX\,J1730, and RX\,J1803. The spectra have been smoothed
with a 3-point boxcar.}
\end{figure*}

\section{Observations: Spectroscopy}
Identification spectroscopy of the target sample and time-resolved
follow-up spectroscopy of newly identified CVs was obtained with four
different telescopes. A brief account of the used instrumentation and
the data reduction procedures is given below.

\subsection{Isaac Newton Telescope identification spectroscopy} 
\label{s-idspec}
The initial identification spectroscopy of ROSAT/2MASS CV candidate
targets was carried out with the Intermediate Dispersion Spectrograph
(IDS) on the Isaac Newton Telescope (INT), located at the Observatorio
del Roque de los Muchachos on La Palma during 7 nights in August 2002
and 7 nights in March 2003 (Table\,\ref{t-obslog}).  The spectra were
obtained using the R400V grating on the 235\,mm camera in conjunction
with a 1.6\,\arcsec slit and an 4k pixel wide EEV CCD, resulting in a
spectral coverage of $3900-7900$\,\AA\ at a resolution of
$\simeq3.5$\,\AA\ at the central wavelength. Standard data reduction
(de-biasing, flat-fielding, optimal extraction, wavelength \& flux
calibration) was done using \texttt{Figaro} within the
\texttt{STARLINK} suite and the \texttt{Pamela/Molly} packages.

While the 2002 run was done under nearly perfect conditions, the 2003
observations were seriously affected by bad weather, resulting in a
number of candidates not being observed. A total of 174 targets were
observed, leading to the discovery of 12 new CVs, including
1RXS\,J063631.9+353537, 1RXS\,J070407.9+262501,
1RXS\,J173021.5--055933, and 1RXS\,J180340.0+401214 (henceforth
RXJ\,0636, RXJ\,0704, RX\,J1730, and RX\,J1803), which are discussed
in this paper). The remaining new CVs as well as a list of all ROSAT
source identifications will be published elsewhere.  Coordinates,
X-ray, optical and infrared properties of the four systems are given
in Table\,\ref{t-targets}, finding charts are shown in
Fig.\,\ref{f-fc}. The INT/IDS identification spectra of RX\,J0636,
RX\,J0704, RX\,J1730 and RX\,J1803 all display Balmer and helium
emission lines (Fig.\,\ref{f-idspec}), identifying them as CVs. The
moderate to strong \Line{He}{II}{4686} emission detected in their
spectra indicates the presence of high-energy photons in these
systems, typically a signature of magnetic CVs, high-mass-transfer
novalike variables and post-novae. RX\,J0636, RX\,J0704 and RX\,J1730
have an equivalent width ratio \Line{He}{II}{4686}/H$\beta>0.5$
(Table\,\ref{t-targets}), which strongly suggests a magnetic nature of
these systems. All three systems were confirmed to be magnetic CVs
indeed by the detection of coherent short-period optical variability,
the hallmark of IPs. RX\,J1803 has a relatively low
\Line{He}{II}{4686}/H$\beta$ ratio for being a magnetic system, but
our photometry clearly identifies this object as an IP as well.

\subsection{Isaac Newton Telescope follow-up spectroscopy}  
Time-resolved follow-up spectroscopy of RX\,J1730 and RX\,J1803 was
obtained at the INT in April/May 2003 using the R632V grating on the
235\,mm camera with a 1.5\,\arcsec slit, covering the range
$4640-6980$\,\AA\ with a spectral resolution of $\simeq2$\,\AA\
(Table\,\ref{t-obslog}). Arc lamp calibration exposures were
interleaved after every three object spectra to account for the
flexure of the instrument. The follow-up spectra were processed in a
standard fashion using the long-slit reduction packages within
\textsc{iraf}\footnote{\textsc{iraf} is distributed by the National
Optical Astronomy Observatories.}. The dispersion solutions were
calculated by fitting a low-order polynomial to the arc data, and the
pixel-to-wavelength relation for the target spectra were obtained from
interpolating between the bracketing arc lamp exposures.

\subsection{Calar Alto 2.2\,m}
Additional time-resolved spectroscopy of RXJ\,1730, RXJ\,1803, and
RX\,J0704 was obtained at the Calar Alto 2.2\,m telescope in
April--June 2003 and in February 2005 using the Calar Alto Faint
Object Spectrograph (CAFOS) equipped with the standard
$2\mathrm{k}\times2\mathrm{k}$ pixel SITe CCD and the G-100
grating. Using a 1.2\,\arcsec slit the observations covered the range
$4240-8300$\,\AA\ at a resolution of $\sim4.1$\,\AA\
(Table\,\ref{t-obslog}). Regular arc lamp exposures were taken to
ensure an accurate wavelength calibration. The reduction of the CAFOS
data was carried out using \textsc{Figaro, Pamela \& Molly} as
described in Sect.\,\ref{s-idspec}.

\subsection{Magellan}
RXJ\,1730 was also observed with the B\&C spectrograph on the 6.5\,m
Magellan-Clay telescope at Las Campanas Observatory in June 2003 under
excellent seeing conditions (0.5\arcsec--0.8\arcsec). The 600 grating blazed
at 5000\AA\ in conjunction with a 0.85\arcsec long-slit provided a
wavelength coverage of 3890--7075\AA\ at 1.55\,\AA/pixel on a
$2048\times515$ Marconi CCD detector. The 58 target exposures
(Table\,\ref{t-obslog}) were reduced and calibrated using the recipe
described in Sect.\,\ref{s-idspec} using regular arc lamp exposures
and nightly observations of the spectrophotometric flux standard
LTT9239.

\subsection{William Herschel Telescope}
Time-resolved spectroscopy of RXJ\,0636 was carried out with the 4\,m
William Herschel Telescope (WHT) on La Palma in January 2005, using the
ISIS double-arm spectrograph with the R600B grating and a 4k$\times$2k
pixel EEV detector in the blue and the R316R grating and a
4.5k$\times$2k pixel Marconi detector in the red. Both arms were
operated with a 1.2\arcsec slit, resulting in a spectral resolution of
$\simeq0.9$\,\AA\ covering the ranges $3600-5000$\,\AA\ and
$6100-8900$\,\AA. Arc lamp and flat-field exposures were taken at
regular intervals to ensure an accurate wavelength calibration and
correct for CCD fringing in the red arm.  The reduction of the WHT
spectroscopy was done using \textsc{Figaro, Pamela \& Molly} as
described in Sect.\,\ref{s-idspec}.

\begin{table*}
\caption{\label{t-targets} X-ray, optical and infrared properties of
the four new CVs.}  \setlength{\tabcolsep}{1.1ex}
\begin{tabular}{lccrrrrrrcccccc}
\hline
System    & RA & Dec & 
 \multicolumn{1}{c}{PSPC} & \multicolumn{1}{c}{HR1} & \multicolumn{1}{c}{HR2} &
 \Ion{He}{II} & H$\beta$ & H$\alpha$ & 
 $B$ & $R$ & $J$ & $H$ & $K$  \\
 & (2000) & (2000) & \multicolumn{1}{c}{$\mathrm{10^{-2}cts\,s^{-1}}$} & & & 
  \AA & \AA & \AA & 
 mag & mag & mag & mag & mag \\
\hline
1RXS\,J063631.9+353537 & 06 36 32.55 & +35 35 43.3 &
 $5.0\pm1.4$ & $0.71\pm0.25$ & $0.33\pm0.29$ &
 8 & 16 & 47 & 
 15.7 & 15.9 & $15.4$ & $15.2$ & $15.1$ \\
1RXS\,J070407.9+262501 & 07 04 08.67 & +26 25 10.9 & 
 $19.4\pm3.3$ & $-0.40\pm0.14$ & $0.55\pm0.26$ & 
 9 & 11 & 24 & 
 16.7 & 16.3 & $16.4$ & $16.4$ & $16.1$ \\
1RXS\,J173021.5--055933 & 17 30 21.90 & $-$05 59 32.1 &
 $58.2\pm4.1$ & $0.86\pm0.03$ & $-0.05\pm0.07$ & 
 8 & 4 & 8 & 
 16.3 & 15.4 & 14.4 & 13.9 & 13.6 \\
1RXS\,J180340.0+401214 & 18 03 39.67 & +40 12 20.6 &
 $17.5\pm1.5$ & $-0.18\pm0.08$ & $-0.02\pm0.12$ &
 18 & 66 & 52 & 
 18.0 & 17.1 & 15.2 & 14.8 & 14.4 \\
\hline
\end{tabular}

\parbox{\textwidth}{Notes. The coordinates and the $B$ and $R$
  magnitudes were taken from the USNO-B catalog \citep{monetetal03-1},
  the ROSAT PSPC count rates and hardness ratios HR1 and HR2 have been
  obtained from the ROSAT All Sky Survey Bright Source Catalogue
  \citep{vogesetal99-1}, the H$\alpha$, H$\beta$ and \Ion{He}{II}
  equivalent widths were measured from the INT/IDS identification
  spectra (Fig.\,\ref{f-idspec}) using the \texttt{integrate/line}
  task in \texttt{MIDAS}, and the $J$, $H$, and $K$ magnitudes were
  taken from the 2MASS All Sky Survey.}
\end{table*}

\begin{figure*}
\begin{center}
\includegraphics[width=6cm]{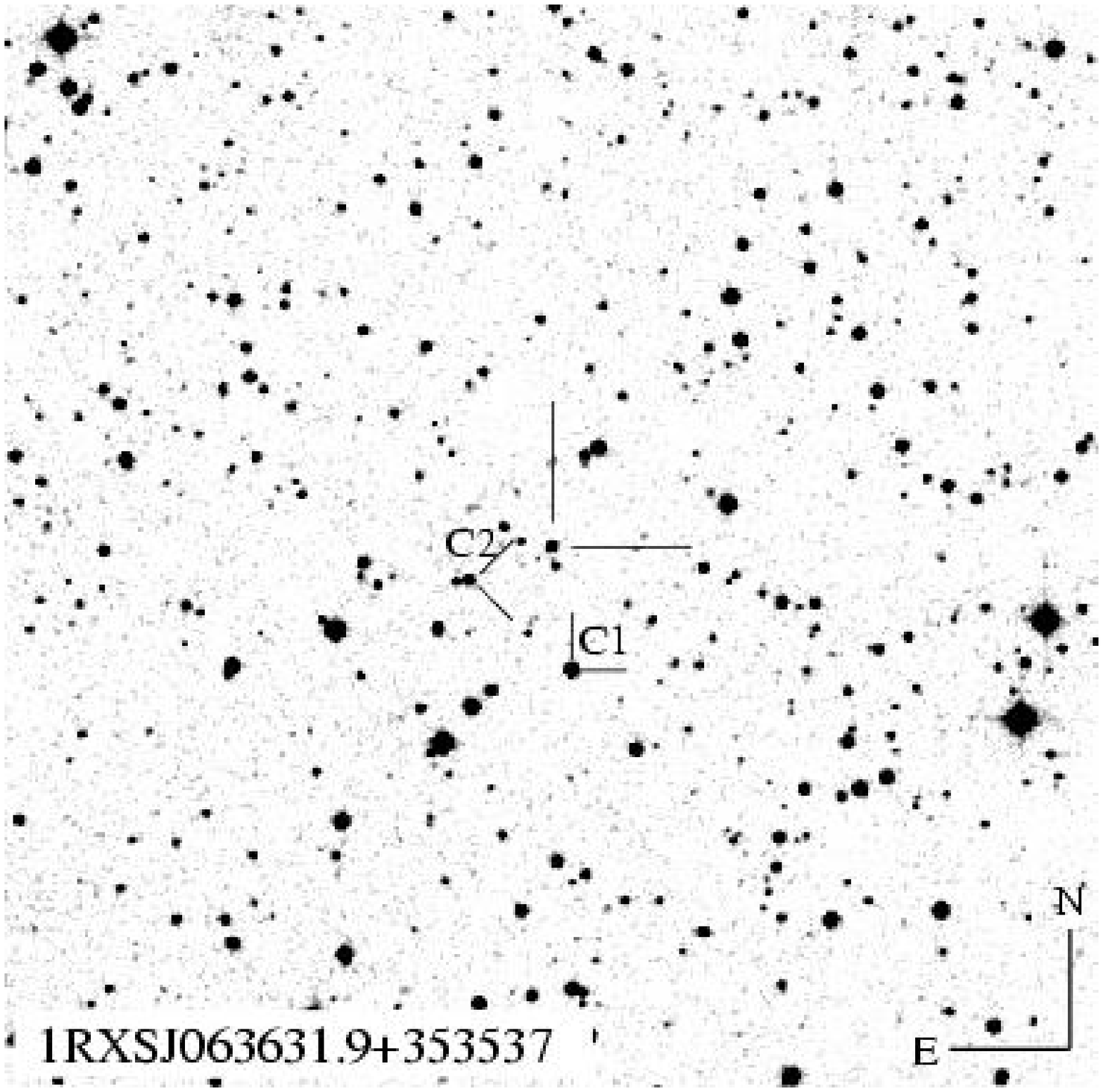}
\hspace*{1cm}
\includegraphics[width=6cm]{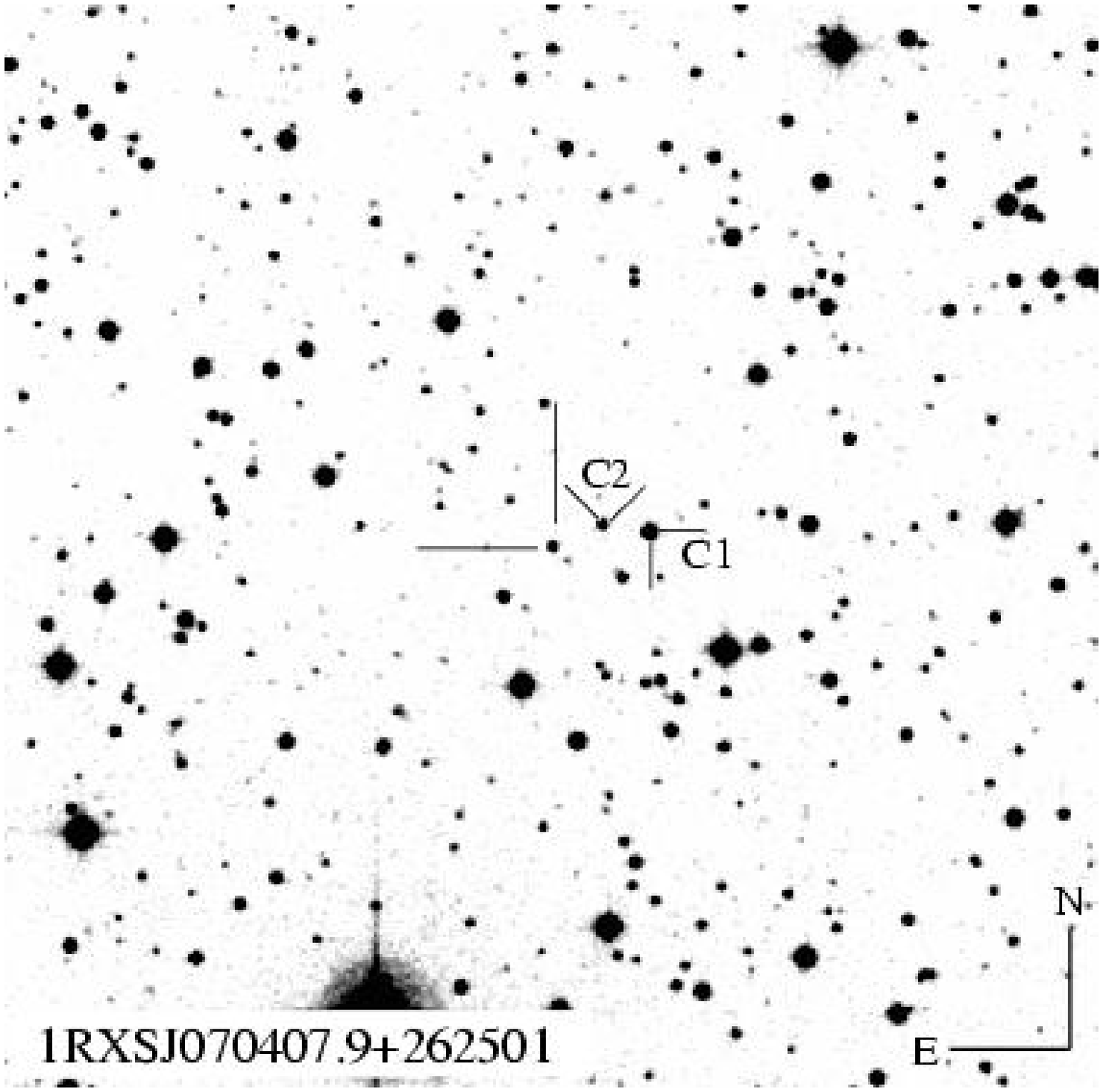}
\end{center}

\begin{center}
\includegraphics[width=6cm]{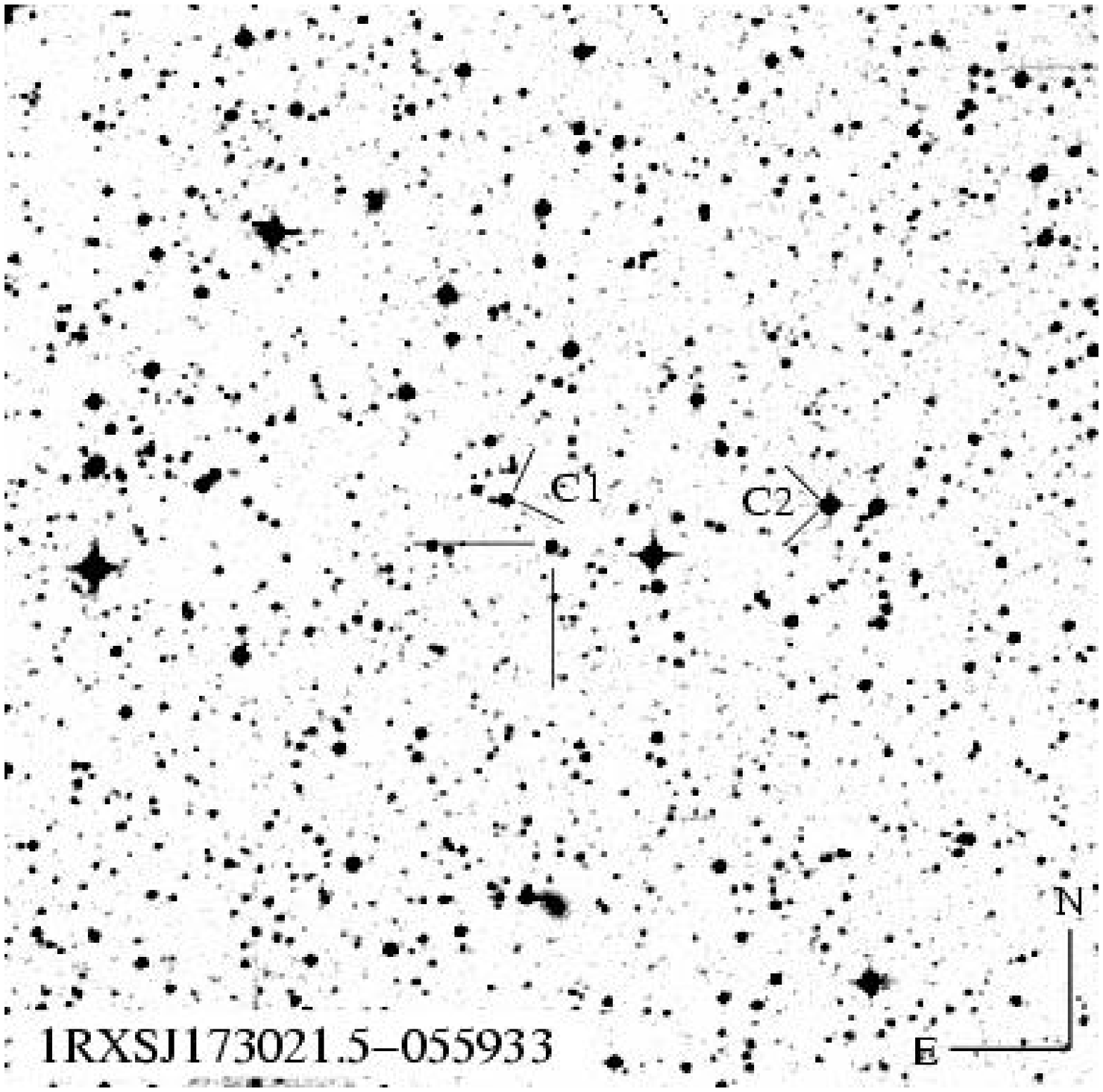}
\hspace*{1cm}
\includegraphics[width=6cm]{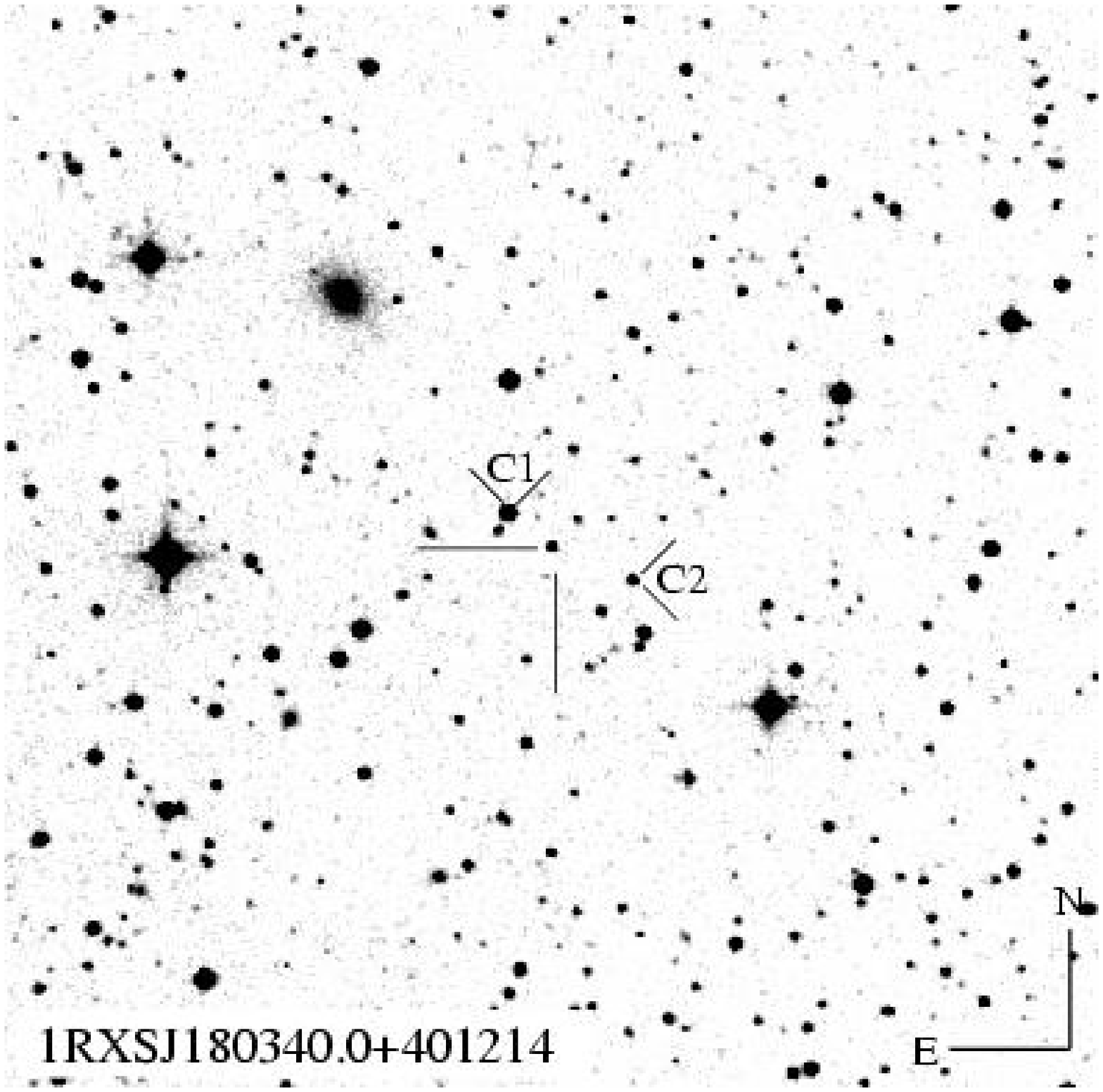}
\end{center}

\caption{\label{f-fc} $10\arcmin\times10\arcmin$ finding charts of
RX\,J0636, RX\,J0704, RX\,J1730, and RX\,J1803 obtained from the
Digitized Sky Survey~2.  Primary and secondary comparison stars for
the CCD photometry are labelled `C1' and `C2', see
Table\,\ref{t-compstars} for details.}
\end{figure*}

\section{Observations: Photometry}
Whereas the INT/IDS spectroscopy suggested a possible magnetic CV
nature for all four systems (Sect.\,\ref{s-idspec}), the
identification and confirmation of the IP nature of RX\,J0636,
RX\,J0704, RX\,J1730 and RX\,J1803 has been derived from CCD time
series obtained at five different telescopes. For all photometric
data, instumental magnitudes of the target and two comparison stars
(labelled `C1' and `C2' in Fig.\,\ref{f-fc}) were extracted from the
images, and differential light curves computed with respect to the
comparison star `C1'. Constant brightness of the primary comparison
star and changes in the observing conditions were checked by
inspecting the differential light curves `C1'--`C2'. The details on
the adopted comparison stars are given in Table\,\ref{t-compstars}.  A
brief description of the instrumentation at the different telescopes
and the data reduction procedures is given below.

\subsection{\label{s-kryoneri}Kryoneri}
Filterless CCD photometry of RX\,J0636, RX\,J0704, and RX\,J1730 was
obtained at Kryoneri Observatory with the 1.2\,m telescope using a
Photometrics SI-502 $516\times516$ pixel camera
(Table\,\ref{t-obslog}). The images were de-biased,
dark-current-subtracted, and flat-fielded within \textsc{MIDAS} and
aperture photometry was performed using  \textsc{Sextractor}
\citep{bertin+arnouts96-1}. A full account of the employed reduction
pipeline is given by \citet{gaensickeetal04-1}.

\subsection{Jacobus Kapteyn Telescope}
The 1.0\,m Jacobus Kapteyn Telescope (JKT) on La Palma was used in May
2003 to obtain $R$-band photometry of RX\,J1730 and $V$-band
photometry of RX\,J1803 using a SITe $2\mathrm{k}\times2\mathrm{k}$
pixel CCD (Table\,\ref{t-obslog}). The reduction was carried out as
described in Sect.\,\ref{s-kryoneri}.

\subsection{Isaac Newton Telescope}
The Wide Field Camera (WFC), an array of 4 EEV
$2\mathrm{k}\times4\mathrm{k}$ pixel CCDs, was used on the INT in
January 2005 to obtain filterless photometry of RX\,J0704. The
read-out time of the WFC is 42\,s, resulting in a somewhat increased
granularity of the observed light curves. The data were processed in an
analogous way as described in Sect.\,\ref{s-kryoneri}.

\subsection{Telescope Nazionale Galileo}
We obtained $g'$ CCD photometry of RX\,J0704 on the 3.6\,m Telescopio
Nazionale Galileo (TNG), located at the Observatorio del Roque de los
Muchachos on La Palma, in January 2005 (Table\,\ref{t-obslog}). We used
DOLORES equipped with a Loral $2\mathrm{k}\times2\mathrm{k}$ pixel
CCD. The reduction was carried out as described in
Sect.\,\ref{s-kryoneri}.

\subsection{Optical Ground Station}
The 1\,m Optical Ground Station (OGS) located at the Observatorio del
Teide on Tenerife was used in July 2003 to obtain filterless and $V$-,
$R$-, and $I$-band photometry of RX\,J1730. The telescope was equipped
with a Thomson $1\mathrm{k}\times1\mathrm{k}$ pixel CCD camera. The
images were bias and flat-field corrected and aligned within
\texttt{IRAF}. Instrumental magnitudes of the CV and the comparison
star `C1' were then extracted using the point spread function (PSF)
photometry tasks package within \texttt{IRAF}.

\begin{table}
\caption{\label{t-compstars} Comparison stars used for the
  differential CCD photometry.}
\begin{tabular}{rlcc}
\hline
System    & USNO reference & $B$  & $R$   \\
\hline
RX\,J0636 & 1200\_04969236 & 15.7 & 15.9  \\ 
    `C1'  & 1200\_04968961 & 15.7 & 14.5  \\
    `C2'  & 1200\_04968814 & 17.3 & 16.0  \\
RX\,J0704 & 1125\_04825852 & 16.3 & 16.2  \\
    `C1'  & 1125\_04824886 & 16.1 & 14.2  \\
    `C2'  & 1125\_04825384 & 18.0 & 16.2  \\
RX\,J1730 & 0825\_10606993 & 16.5 & 15.8  \\
    `C1'  & 0825\_10607901 & 16.9 & 15.5  \\
    `C2'  & 0825\_10601801 & 15.8 & 13.7  \\
RX\,J1803 & 1275\_09738647 & 17.3 & 17.2  \\
    `C1'  & 1275\_09739056 & 15.1 & 14.1  \\
    `C2'  & 1275\_09737883 & 17.9 & 16.7  \\
\hline
\end{tabular}
\end{table}

\section{Analysis and Results}

\subsection{1RXS\,J063631.9+353537} 
The INT/IDS identification spectrum of RX\,J0636
(Fig.\,\ref{f-idspec}) is characterised by narrow Balmer and He
emission lines superimposed on a blue continuum.  The detection of
relatively strong \Line{He}{II}{4686} suggests a potential magnetic CV
nature for this system. No obvious signature of the donor star is
detected in the spectrum.

\begin{figure}
\includegraphics[angle=-90,width=\columnwidth]{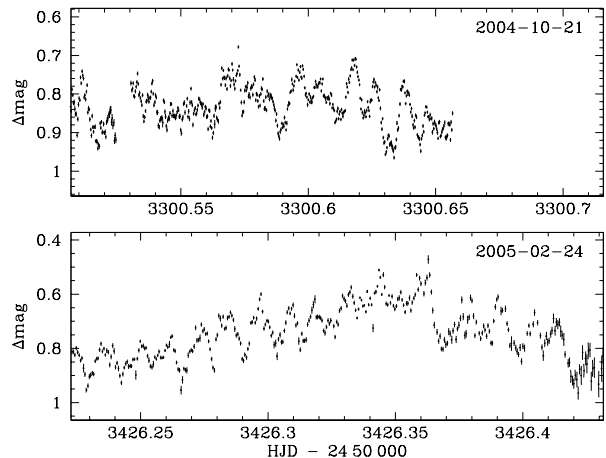}
\caption{\label{f-lc0636} Sample light curves of RX\,J0636 obtained
with the 1.2\,m Kryoneri telescope.}
\end{figure}

\begin{figure}
\includegraphics[angle=-90,width=\columnwidth]{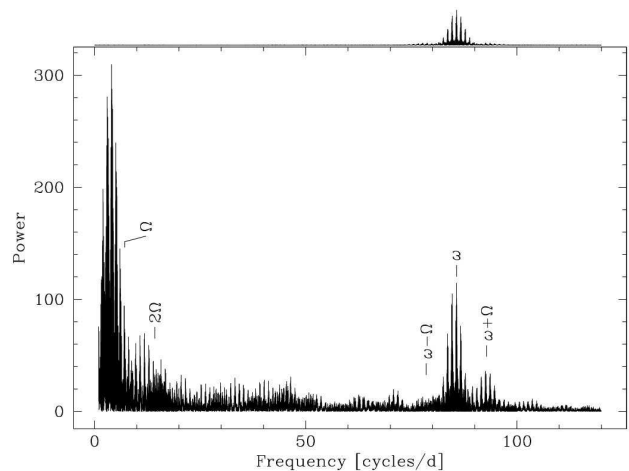}
\caption{\label{f-scargle0636}Scargle periodogram computed from the
  combined photometric data of RX\,J0636 (Table\,\ref{t-obslog}). We
  tentatively assign the peak detected at 85.7\,\id\ (1008\,s) to the
  white dwarf spin frequency $\omega$ and the peak at 92.9\,\id\
  (930\,s) to the orbital side band $\omega+\Omega$. The location of
  the $\omega-\Omega$ side band is indicated. We caution, however,
  that the data does not allow an unambiguous identification, and that
  the true spin and side band signals may be swapped.  No significant
  power is detected at the orbital period or its harmonic. The
  strong signal near 4\,\id\ appears only in the 2005 data, and its
  nature is not clear. Shown on top of the Figure is the window
  function of the data set, shifted to $\omega$.}
\end{figure}

\begin{figure}
\includegraphics[angle=-90,width=\columnwidth]{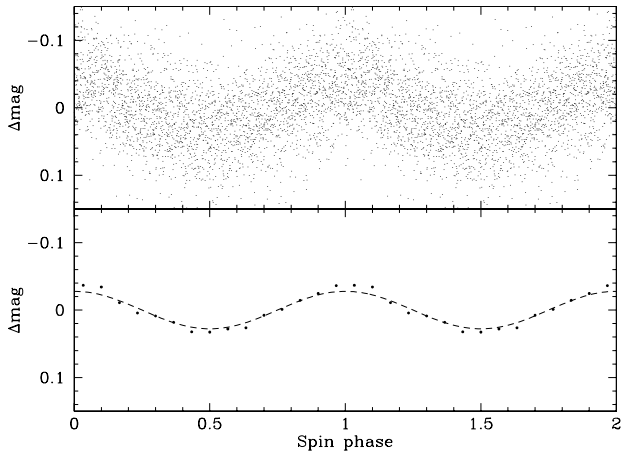}
\caption{\label{f-spinfold0636} Spin-folded light curves of RX\,J0636,
  assuming $\Pspin=1008.3408$\,s. Top panel: all individual data
  points (Table\,\ref{t-obslog}). Variability on longer time scales
  has been removed from each individual night by subtracting the light
  curve smoothed by a 30-point box car.  Bottom panel: Data binned
  into 30 phase slots. Plotted as dashed line is a sine fit to the
  binned \& folded data.}
\end{figure}

\subsubsection{Photometry}
The CCD photometry of RX\,J0636 obtained at Kryoneri observatory in
October 2004 and February/March 2005 (Table\,\ref{t-obslog},
Fig.\,\ref{f-lc0636}) displays substantial variability on time scales
of $\sim15$\,min, superimposed on longer-term
modulations. Figure\,\ref{f-scargle0636} shows a Scargle
(\citeyear{scargle82-1}) periodogram computed from the combined
photometric data, using the \texttt{MIDAS/TSA} context. The suspected
presence of a coherent short-period modulation is confirmed by the
detection of relatively strong signal at $\simeq85.7$\,\id, with a
1-day alias of similar strength at $\simeq84.7$\,\id. A second cluster
of signals peaks at $\simeq92.9$\,\id\ and $\simeq91.9$\,\id. The
periods corresponding to the highest peaks of each cluster of signals
are 1008.3408(19)s and 930.5829(40)s, respectively, where the errors
have been determined from fitting a sine wave to the photometric
data. In order to facilitate an improvement in the accuracy of the
spin period, we have fitted a sine wave with the period fixed to the
best-value determined from the entire data set to each individual
night of photometry, and report in Table\,\ref{t-hjdzero} the times of
spin maxima determined from these fits. Folding the photometric data
over the 1008s period results in a quasi-sinusoidal modulation with an
average amplitude of 0.03\,mag (Fig.\,\ref{f-spinfold0636}). Given the
possible magnetic nature of RX\,J0636 suggested by its spectroscopic
appearance, we tentatively interpret these two signals as the white
dwarf spin frequency (period) $\omega$ (\Pspin) and an orbital
sideband (beat) frequency $\omega\pm\Omega$, with $\Omega$ the orbital
frequency, as the detection of these signals is the hallmark of the
magnetic IPs. The optical photometry alone does not provide sufficient
information to unambiguously identify the two signals, as both spin
and beat dominated IPs are known. If the two photometric signals were
indeed $\omega$ and $\omega\pm\Omega$, RX\,J0636 is expected to have
an orbital period of $\simeq201$\,min. No significant power is
detected in the periodogram near the corresponding frequency of
$\simeq7.15$\,\id\ (Fig.\,\ref{f-scargle0636}). Puzzling is the strong
signal near $\simeq4$\,\id\ ($\simeq360$\,min), which is detected in
all the 2005 data, but absent in the 2004 data. The period of this
photometric modulation is clearly much longer than the orbital period
estimated from the beat period, and independently from the
time-resolved spectroscopy discussed below. Additional photometric
observations will be necessary to confirm that this 360\,min
periodicity is a coherent and repeating characteristic of RX\,J1803.

\begin{figure}
\includegraphics[angle=-90,width=\columnwidth]{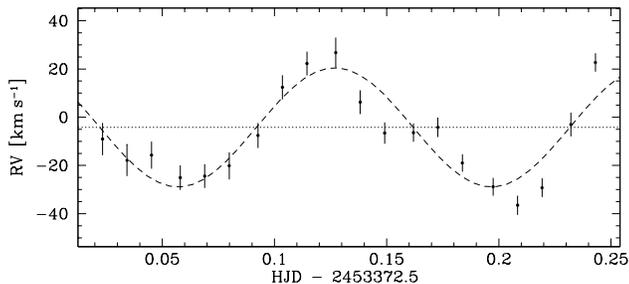}
\caption{\label{f-rv0636} Radial velocity (RV) variation of H$\alpha$
  measured from the time-resolved WHT spectroscopy of RX\,J0636
  (Table\,\ref{t-obslog}). Plotted as dashed line is a sine fit to the
  data, resulting in a period of 201(8)\,min. The dotted line 
  indicates the systemic velocity determined from the sine fit,
  $\gamma=-4(2)$\,\kms. }
\end{figure}

\begin{table}
\begin{minipage}{\columnwidth}
\caption{\label{t-hjdzero} HJD of the spin maxima (RX\,J0636,
  RX\,J0704, RX\,J1803) or spin minima (RX\,J1730), obtained from
  sine-fits to each individual night of photometric data.}
\setlength{\tabcolsep}{1.1ex}
\begin{tabular}{lr@{\hspace{5ex}}lr}
\hline
System & HJD & System & HJD \\
\hline
\textbf{RX\,J0636} & 2453299.564084 & \textbf{RX\,J1730} & 2452769.552687\\
                   & 2453300.499631 &                    & 2452780.528206\\
                   & 2453301.491248 &                    & 2452828.411952\\
                   & 2453302.470448 &                    & 2452829.411925\\
                   & 2453423.366908 &                    & 2452830.397171\\
                   & 2453425.234453 &                    & 2452830.397133\\
                   & 2453426.214890 &                    & 2452836.385213\\
                   & 2453433.240185 &                    & 2452837.395621\\
\noalign{\smallskip}
\textbf{RX\,J0704} & 2453346.611013 & \textbf{RXJ\,1803} & 2452781.436952\\
                   & 2453372.526838 &                    & 2452867.257081\\
                   & 2453376.482737 &                    & 2452868.260431\\
                   &                &                    & 2452869.245696\\
                   &                &                    & 2452870.248852\\
                   &                &                    & 2452871.252155\\
\hline
\end{tabular}
\end{minipage}
\end{table}

\subsubsection{Time-resolved spectroscopy}
In order to test our hypothesis of RX\,J0636 being a CV with an
orbital period near 201\,min we have obtained 5.3\,h time-resolved
spectroscopy at the WHT (Table\,\ref{t-obslog}). Radial velocity
measurements were carried out using a single Gaussian fit to
H$\alpha$, the strongest emission line, fixing the full-width at
half-maximum (FWHM) to 200\,\kms. The resulting radial velocity curve
clearly shows a quasi-sinusoidal modulation with a amplitude of
$\simeq25$\,\kms (Fig.\,\ref{f-rv0636}). A Scargle periodogram
computed from the radial velocity measurements shows a broad peak
centred at 7.3\,\id, corresponding to 197\,min. A sine fit to the
radial velocities results in $P=201(8)$\,min, which is consistent with
the orbital period estimate obtained from the photometry.

Based on the currently available spectroscopic and photometric
observations, we consider RX\,J0636 to be an IP, with an orbital period of
$\Porb\simeq201$\,min and a white dwarf spin period of either
$\Pspin=1008.3048$\,s or $\Pspin=930.5829$\,s.

\subsection{1RXS\,J070407.9+26250}
The INT/IDS identification spectrum or RX\,J0704
(Fig.\,\ref{f-idspec}) is overall similar to that of RX\,J0636, 
with the difference of displaying somewhat broader lines. No signature
of the secondary star is detected.

\begin{figure}
\includegraphics[angle=-90,width=\columnwidth]{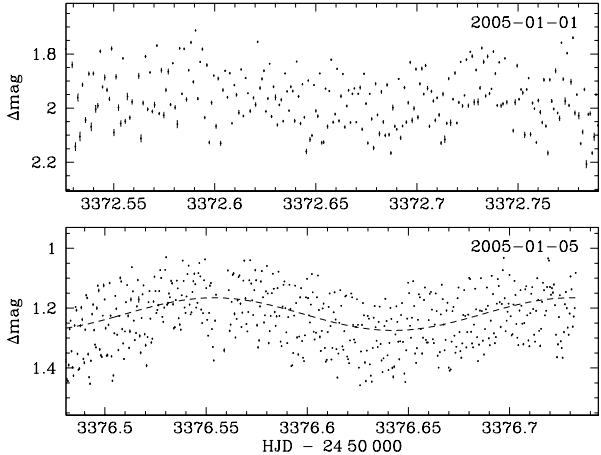}
\caption{\label{f-lc0704} Sample light curves of RX\,J0704 obtained
  with the INT/WFC (top panel) and the TNG/DOLORES (bottom panel). The
  light curves are dominated by a modulation at twice the white dwarf
  spin period $\Pspin=480.708$\,s, and a longer-period modulation
  with $P\simeq4$\,h, which we tentatively identify as the orbital
  period. A sine fit to the TNG data gives $\Porb=254.7(1.1)$\,min
  (dashed curve).}
\end{figure}

\begin{figure}
\includegraphics[angle=-90,width=\columnwidth]{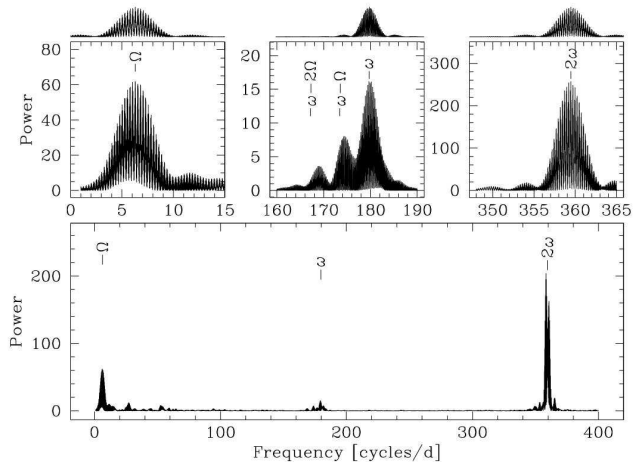}
\caption{\label{f-scargle0704} Scargle periodogram computed from the
  combined photometric data of RX\,J0704
  (Table\,\ref{t-obslog}). Bottom panel: full frequency range. Middle
  panels, left to right: enlargement around the orbital, the white
  dwarf spin, and twice the white dwarf spin frequencies. Top panels:
  the window function calculated from the temporal sampling of the
  data, shifted to the assumed orbital ($\Omega$), spin ($\omega$) ,
  and first harmonic of the spin $2\omega$ frequencies. The
  corresponding periods are $\simeq250$\,min, 480.708\,s, and
  240.354\,s, respectively.}
\end{figure}

\begin{figure}
\includegraphics[angle=-90,width=\columnwidth]{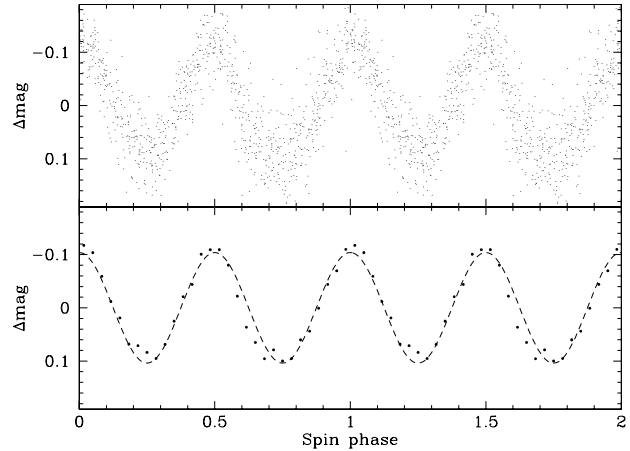}
\caption{\label{f-spinfold0704} Spin-folded light curves of RX\,J0704
  using $\Pspin=480.708$\,s. Top panel: all individual data points
  (Table\,\ref{t-obslog}). The putative orbital modulation has been
  removed prior to phase-folding by means of a 30-point boxcar. Bottom
  panel: Data binned into 30 phase slots. Plotted as dashed line is a
  sine fit to the binned \& folded data.}
\end{figure}

\subsubsection{Photometry}
The first brief photometric data set obtained at Kryoneri
(Table\,\ref{t-obslog}) displayed substantial short-term variability
with a peak-to-peak amplitude of $\sim0.3$\,mag. A Scargle periodogram
computed from those data contained a strong signal at $\sim360$\,\id,
identifying RX\,J0704 as an IP candidate with a rather short white
dwarf spin period. Figure\,\ref{f-lc0704} shows the more extensive
photometry obtained at the INT and TNG (Table\,\ref{t-obslog}), which
revealed a modulation with a period of $\sim4$\,h in addition to the
spin variability. The strongest peak in a Scargle periodogram
calculated from the entire data set is found at a period of 4\,min
(Fig.\,\ref{f-scargle0704}), consistent with the earlier Kryoneri
result. However, the larger data set shows also power at 480\,s, which
we subsequently interpret as the white dwarf spin. The 240\,s period
is then the first harmonic of the spin, most likely due to the
changing aspect of two accreting regions on/close to the white
dwarf (i.e. the accretion funnels and white dwarf pole caps, see e.g.
\citealt{hellier95-1}). The periodogram around \Pspin\ and
2\,\Pspin\ is severely plagued by aliases, caused by the sparse data
coverage over the base line of one month. The 240\,s signal displays
two nearly equally strong aliases, their periods determined from sine
fits to the photometric data are $240.3540(22)$\,s and
$240.5329(22)$\,s, and we adopt $\Pspin=480.7080(44)$\,s as the
putative white dwarf spin.  Table\,\ref{t-hjdzero} lists the times of
spin maxima for each night of photmetry, determined from fitting a
sine-wave to the individual data sets (where the period was fixed to
the value determined from the entire data set). As the two maxima per
spin cycle are almost equal, there is the chance of a 0.5~cycle
mismatch.  We have detrended the $\sim4$\,h modulation from the
photometric data by means of a 30-point boxcar, and show in
Fig.\,\ref{f-spinfold0704} the detrended data folded over
$\Pspin$. The spin light curve is quasi-sinusoidal with two nearly
equal maxima per spin cycle, explaining the dominant power at
2\,$\Pspin$. A sine fit to the folded data reveals some small
differences in the pulse durations, which is the reason for the
detection of power at the actual $\Pspin$.

The $\sim4$\,h modulation (especially conspicuous in the TNG light
curve, Fig.\,\ref{f-lc0704} bottom panel), is reflected by the
second strongest signal in the periodogram of the combined photometric
data (Fig.\,\ref{f-scargle0704}). Formally this implies
$P=228.73(5)$\,min, but the severe aliasing prevents a unique
idenfication of the true period. A sine fit to the TNG data alone
results in $P=254.7(1.1)$\,min. The close-up of the periodogram around
the white dwarf spin frequency $\omega$ (Fig.\,\ref{f-scargle0704},
top middle panel) shows substantial power excess with respect to the
window function in the form of two clusters of aliases at frequencies
below $\omega$. The separation between these signals and $\omega$ is
$5.68\,\id$ and $11.24\,\id$. If interpreted as the beat frequencies
$\omega-\Omega$ and $\omega-2\Omega$, which are commonly detected in
IPs, these signals imply an orbital period of $\simeq250$\,min. This
falls well within the alias structure of the low-frequency signal,
and is consistent with the period derived from the best (TNG)
photometric data set.

\begin{figure}
\includegraphics[angle=-90,width=\columnwidth]{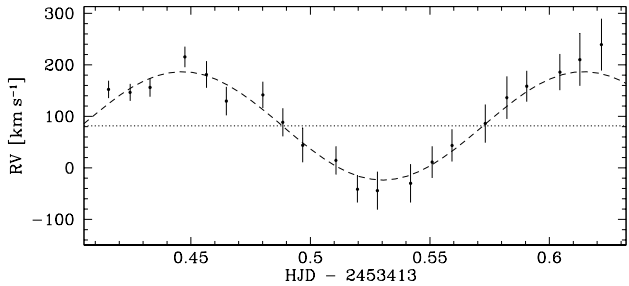}
\caption{\label{f-rv0704} Radial velocity (RV) variation of the
  H$\alpha$ wings measured from the time-resolved Calar Alto
  spectroscopy of RX\,J0704 (Table\,\ref{t-obslog}). Plotted as dashed
  line is a sine fit to the data, resulting in a period of
  242(7)\,min. The dotted line indicates the systemic velocity
  determined from the sine fit, $\gamma=82(5)$\,\kms. }
\end{figure}

\subsubsection{Time-resolved spectroscopy}
A single 5.6\,h run of time-resolved spectroscopy of RX\,J0704 was
obtained at Calar Alto in February 2005 (Table\,\ref{t-obslog}), with
the aim to derive a dynamical estimate of the orbital period. The
radial velocity variation of H$\alpha$ was measured using a double
Gaussian convolution \citep{schneider+young80-2} with a separation of
1000\,\kms\ and a FWHM of 400\,\kms. This H$\alpha$ radial velocity curve 
displays a quasi-sinusoidal modulation with a semi-amplitude of
$\simeq110$\,\kms (Fig.\,\ref{f-rv0704}). The  Scargle periodogram
computed from these data contains a broad peak near 5.7\,\id\
(253\,min). Fitting a sine wave to the radial velocity data gives 
$P=242(7)$\,min, which agrees well with the period estimate derived
from the TNG photometry. 

We conclude on the base of the spectral, photometric, and X-ray
properties of RX\,J0704 that this object is an IP with an orbital
period of $\simeq250\,\mbox{min}$ and a spin period of 480.708\,s. The
strong power at the first harmonic of the spin period suggests that
both accretion regions contribute to a similar extent to the optical
light.

\subsection{1RXS\,J173021.5--055933} 
The identification of RX\,J1730 is the reddest of all four objects
presented here, and displays the lowest emission line equivalent
widths. Noticeable is the large equivalent width of
\Line{He}{II}{4686}, which exceeds that of H$\beta$ by a factor
$\simeq2$. The broad absorption feature near 5200\,\AA\ in
Fig.\,\ref{f-idspec} is an artifact from the data reduction that has
been found in a number of other target stars.

\begin{figure}
\includegraphics[angle=-90,width=\columnwidth]{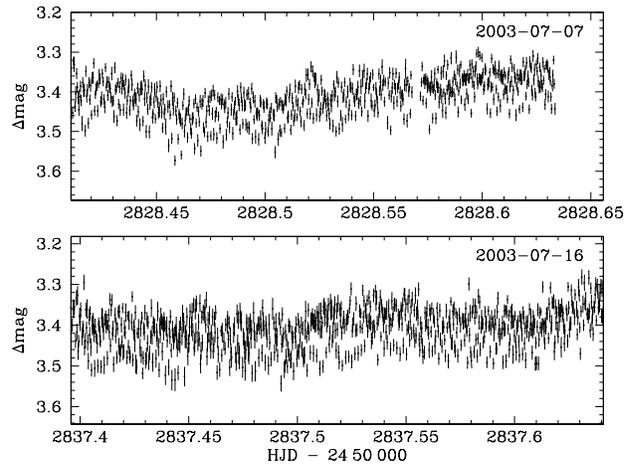}
\caption{\label{f-lc1730}Sample filterless light curves of RX\,J1730
  obtained with the 1\,m OGS.}
\end{figure}

\begin{figure}
\includegraphics[angle=-90,width=\columnwidth]{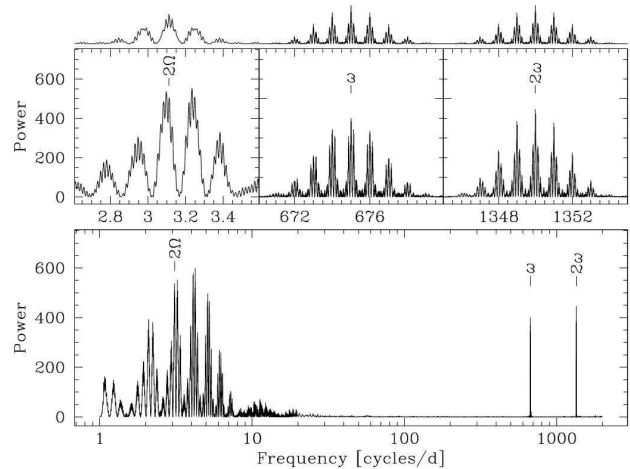}
\caption{\label{f-scargle1730} Scargle periodogram computed from the
  combined photometric data of RX\,J1730
  (Table\,\ref{t-obslog}). Bottom panel: full frequency range, note
  the logarithmic frequency scale. Unambiguous signals are the white
  dwarf spin frequency at $\omega=675$\,\id\
  ($\Pspin=128$\,s). Orbital variability is detected at twice the
  orbital frequency ($\Porb=925.27$\,min), most likely reflecting the
  ellipsoidal modulation from the secondary star. Top panels, left to
  right: enlargement around the orbital, the white dwarf spin, and
  twice the white dwarf spin frequencies. Top panels: the window
  function calculated from the temporal sampling of the data, shifted
  to the assumed $2\Omega$, $\omega$, and $2\omega$.}
\end{figure}

\begin{figure}
\includegraphics[angle=-90,width=\columnwidth]{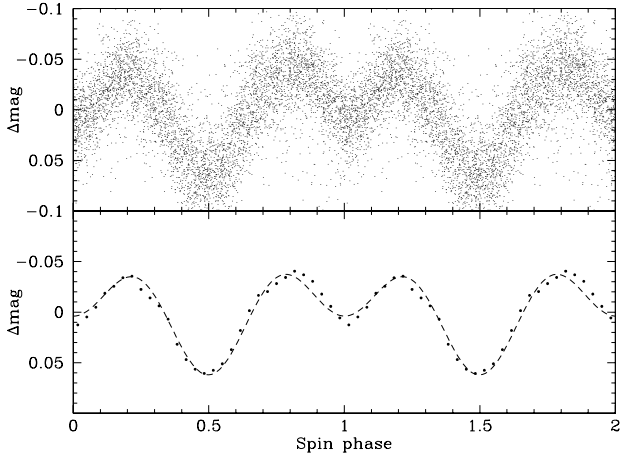}
\caption{\label{f-spinfold1730} The photometric data of RX\,J1730
  folded over the 128\,s spin period. Top panel: all individual data
  points. Bottom panel: data averaged into 30 phase bins. A
  two-frequency fit ($\omega$ and $2\omega$) is shown as dashed line.}
\end{figure}

\subsubsection{Photometry}
The JKT $R$-band photometry of RX\,J1730 obtained in May 2003
(Table\,\ref{t-obslog}) displayed short term variability with an
amplitude of $\simeq0.05$\,mag, much in excess to the photometric
errors of the individual data points. A Scargle periodogram computed
from this data revealed strong signals near 675\,\id\ and
1350\,\id. The latter frequency is undersampled by the data, resulting
in a number of spurious aliased signals. Prompted by this
discovery, high-time resolution photometry of RX\,J1730 was obtained
with the OGS in July 2003 (Table\,\ref{t-obslog}). Sample light curves
of this run are shown in Fig.\,\ref{f-lc1730}. The periodogram computed
from the OGS data displays three distinct clusters of signals centred
on 4.2\,\id, 675\,\id, and 1350\,\id\ (Fig.\,\ref{f-scargle1730}). The
nature of the low-frequency signal will be discussed below together
with the time-resolved spectroscopy.

The high-frequency signals unambiguously identify RX\,J1730 as an
IP. We interpret the detection of the two commensurate signals as the
white dwarf spin frequency and its first harmonic, suggesting that
both accretion regions contribute to the observed photometric
modulation. Combining the OGS and JKT data, we determine
$\Pspin=127.999909(49)$\,s. However, the two neighbouring aliases at
128.002689\,s and 127.997256\,s can not be excluded.
Table\,\ref{t-hjdzero} lists the times of spin minima (which can be
unambiguously identified) for each night of photmetry, determined from
fitting two sine-waves to the individual data sets (where the periods
were fixed to \Pspin\ and 2\Pspin\ as determined from the entire data
set, but phases were left free for both sine waves).  With a spin
period of 128\,s, RX\,J1730 is the second-fastest spinning white dwarf
in an IP, behind AE\,Aqr ($\Pspin=33.08$\,s) and followed by DQ\,Her
($\Pspin=140.1$\,s). The spin-folded light curve
(Fig.\,\ref{f-spinfold1730}) shows two maxima per spin cycle, with
unequal minima. A two-frequency sine fit, with the higher frequency
being fixed as the harmonic of the lower one, results in nearly equal
amplitudes ($\simeq0.03$) and a phase offset of $\sim0.57$, suggesting
nearly diametrically opposite accretion regions on/close to the white
dwarf.

Comparing the amplitudes of the spin signal in the filtered photometry
obtained at the OGS in July 2003 ($I$: 0.03\,mag, $R$: 0.035\,mag,
$V$: 0.057\,mag) suggests that the spectrum of the 128\,sec variabilty
is blue, as expected for the emission from a hot pole
cap/accretion funnel. Additional data in $B$ and $U$ would be
desirable to estimate a colour temperature for the spin modulation.

\begin{figure}
\includegraphics[angle=-90,width=\columnwidth]{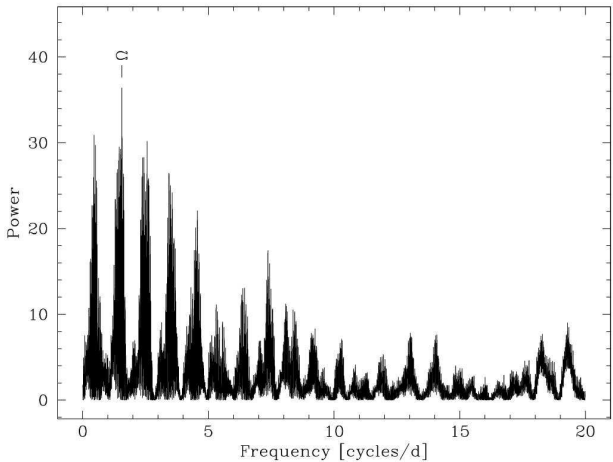}
\caption{\label{f-rvscargle1730} Scargle periodogram computed from the
  radial velocity variation of the \Line{He}{II}{4686} line wings in
  RX\,J1730. The inferred orbital period is $\Porb=925.27$\,min.}
\end{figure}

\begin{figure}
\includegraphics[angle=-90,width=\columnwidth]{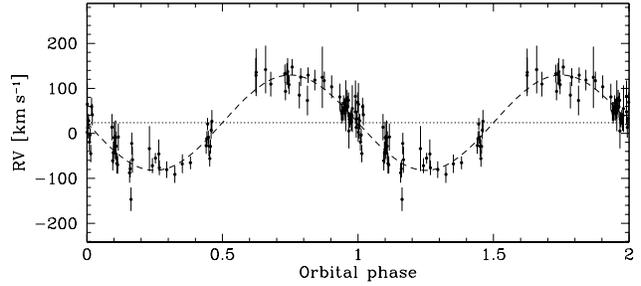}
\caption{\label{f-rvfold1730} The \Line{He}{II}{4686} radial
  velocity (RV) measurements of RX\,J1730 folded over the orbital period of
  925.27(14)\,min. Shown as a dashed line is a sine fit to the folded
  radial velocity data. The systemic velocity determined from the fit
  is $\gamma=22(3)$\,\kms.}
\end{figure}

\subsubsection{Time-resolved spectroscopy}
The main purpose of obtaining time-resolved spectroscopy of RX\,J1730
was to determine the orbital period of the system. The first set of
INT data (April 2003, Table\,\ref{t-obslog}) suggested a rather long
period, prompting additional spectroscopic follow-up at the INT, Calar
Alto, and Magellan. All spectra were binned on the dispersion of the
lowest-resolution data (Calar Alto), corrected for the heliocentric
velocity of the Earth, and continuum-normalised around H$\alpha$ (not
covered by the April 2003 data), H$\beta$ and
\Line{He}{II}{4686}. Radial velocity measurements were carried out by
means of correlating either a single Gaussian or a double Gaussian
\citep{schneider+young80-2}, using a variety of widths (and
separations in the case of the double Gaussian method). The radial
velocity measurements were then subjected to a Scargle periodogram
calculation. Whereas the analysis of the Balmer lines did not provide
any conclusive result, the radial velocities measured from the
\Line{He}{II}{4686} wings using a double-Gaussian convolution with a
$\mathrm{FWHM}=200$\,\kms\ and a separation of 1200\,\kms\ yielded a
well-defined signal at 1.556\,\id\ in the periodogram
(Fig.\,\ref{f-rvscargle1730}). A period of 925.27(14)\,min is derived
from a sine fit to the \Ion{He}{II} radial velocity data, which we
interpret as the orbital period of the
system. Figure\,\ref{f-rvfold1730} shows the \Ion{He}{II} radial
velocities folded over the orbital period. Inspecting the periodogram
computed from the photometry (Fig.\,\ref{f-scargle1730}), it is
apparent that twice the spectroscopic frequency coincides well with
the one of the alias clusters in the low-frequency range, which
suggests that ellipsoidal modulation is responsible for low-frequency
photometric variability.

\begin{figure}
\includegraphics[width=\columnwidth]{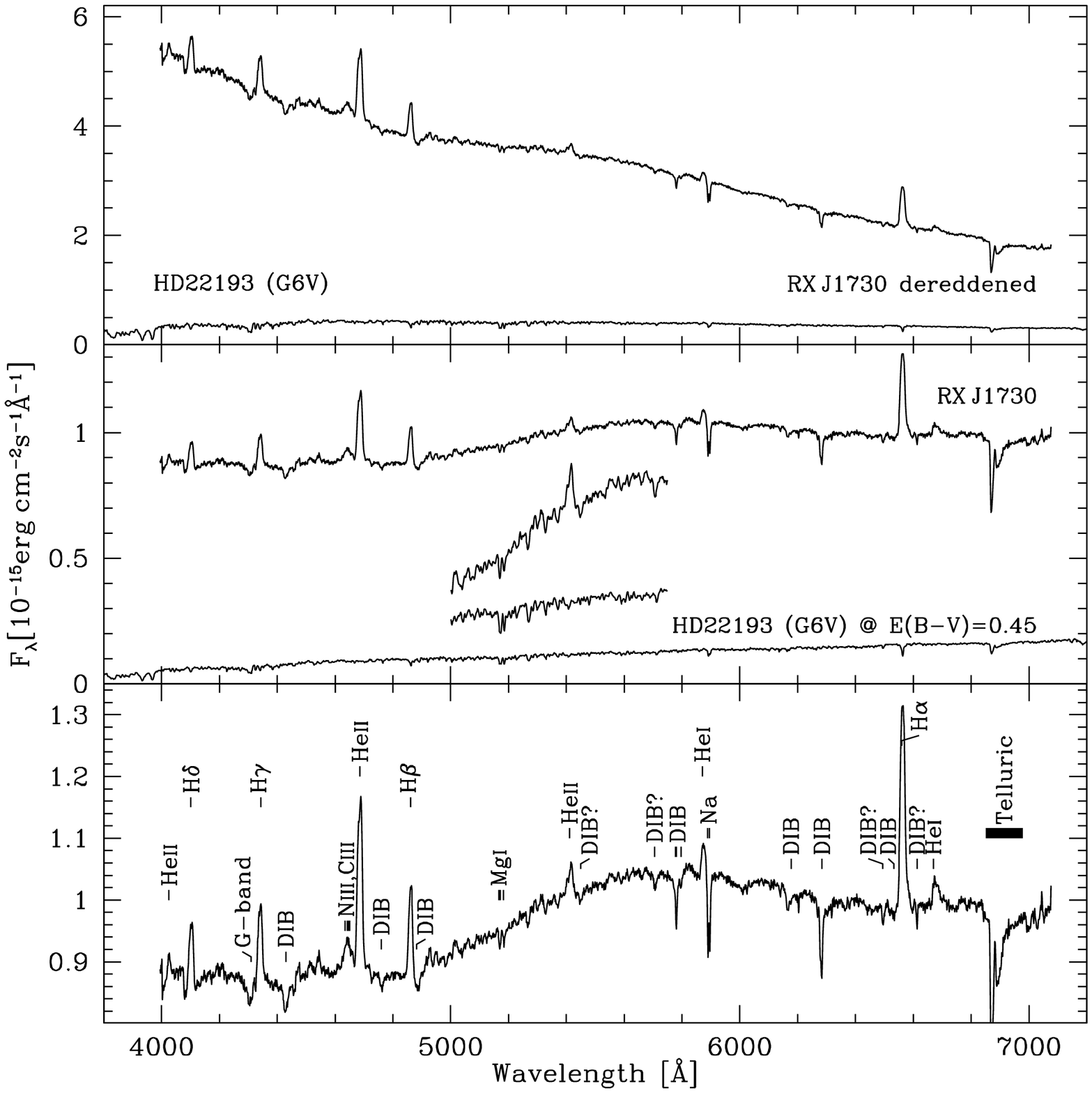}
\caption{\label{f-rxj1730spec} Bottom panel: the average of the
  Magellan spectra of RX\,J1730 obtained in June 2003. The major
  emission and absorption features are identified. The most noticeable
  features that are likely to be associated with the secondary star
  are the \Lines{Mg}{I}{5167,5172,5183} triplet and the $\lambda4310$
  G-band. All the strong absorption features are due to diffuse
  interstellar bands and interstellar sodium absorption. Middle panel:
  the average Magellan spectrum of RX\,J1730 along with a G6V star
  that has been reddened by $E(B-V)=0.45$ and that contributes 15\,\%
  of the observed flux at 6800\,\AA. This main-sequence star spectrum
  approximately reproduces the strength of the observed stellar
  absorption features in RX\,J1730. The inset shows an enlarged view
  of the two spectra, demonstrating that a number of weak absorption
  lines are contained in both spectra. Top panel: the average Magellan
  spectrum dereddened by $E(B-V)=0.45$, along with the (now
  unreddened) G6V star from the middle panel.}
\end{figure}

\subsubsection{The Magellan average spectrum}
Given the long orbital period of RX\,J1730, one might expect to detect
its secondary star at optical wavelengths. In fact, a small
contribution to the optical flux is suggested by the detection of
ellipsoidal modulation in the photometric
data. Figure\,\ref{f-rxj1730spec} shows the average of the
flux-calibrated spectra of RX\,J1730, which were obtained under very
good atmospheric conditions. Whereas the majority of the strong
absorption features are identified with diffuse interstellar bands
(DIBs, \citealt{herbig95-1}) or atomic interstellar absorption
(\Ion{Na}{I}), weak absorption of \Lines{Mg}{I}{5167,5172,5183} and
the $\lambda\lambda\,4310$ G-band are detected. Considering the
strength of the DIBs, it is clear that the spectrum must be
significantly reddened. For the galactic position of RX\,J1730 the
dust maps by \citet{schlegeletal98-1} give an upper limit of
$E(B-V)=0.6$. An independent estimate is obtained using the neutral
hydrogen maps of \citet{dickey+lockman90-1}, which gives
$\Nh=1.66\times10^{21}\,\mathrm{cm^{-2}}$ for the line of sight
towards RX\,J1730, corresponding to an upper limit of $E(B-V)=0.34$
\citep{diplas+savage94-1}. For the following discussion, we will
assume $E(B-V)=0.45$. The weakness of the \Ion{Mg}{I} triplet and the
G-band indicates that the secondary star contributes only a small
fraction of the optical flux. In order to estimate the spectral type
of the secondary star, we have taken mid-K to mid-F star from the
\citet{jacobyetal84-1} library, subjected them to an $E(B-V)=0.45$
reddening, and then scaled them in flux to approximately match the
strength of the \Ion{Mg}{I} triplet and the G-band absorption features
observed in RX\,J1730.  The absence of the broad
$\lambda\lambda\,5000-5200$ absorption band excludes a spectral type
later than $\simeq$K0, and we find that spectral types in the range
G4V to G0V, contributing $\sim15$\% at 6800\,\AA, provide the best
match for the observed line strengths
(Fig.\,\ref{f-rxj1730spec}). From this exercise it appears that the
accretion flow is the dominant source of optical light in RX\,J1730.

We conclude that RX\,J1730 is an IP with an orbital period of 925.27\,min
and a white dwarf spin period of 128.0\,s, which resembles in some
aspects the enigmatic system AE\,Aqr. Differences are that its optical
spectrum is dominated by the accretion flow and not by the secondary
star, and that (so far) none of the flaring episodes characteristic of
AE\,Aqr has been observed in RX\,J1730.

\subsection{1RXS\,J180340.0+401214} 
The IDS identification spectrum of RX\,J1803 (Fig.\,\ref{f-idspec})
shows a typical CV spectrum with a negative Balmer decrement and
relatively weak \Line{He}{II}{4686} emission, suggesting a possible
magnetic nature of the system. Of all four CVs presented in this
paper, RX\,J1803 has the lowest \Line{He}{II}{4686}/H$\beta$ flux
ratio (Table\,\ref{t-targets}).

\begin{figure}
\includegraphics[angle=-90,width=\columnwidth]{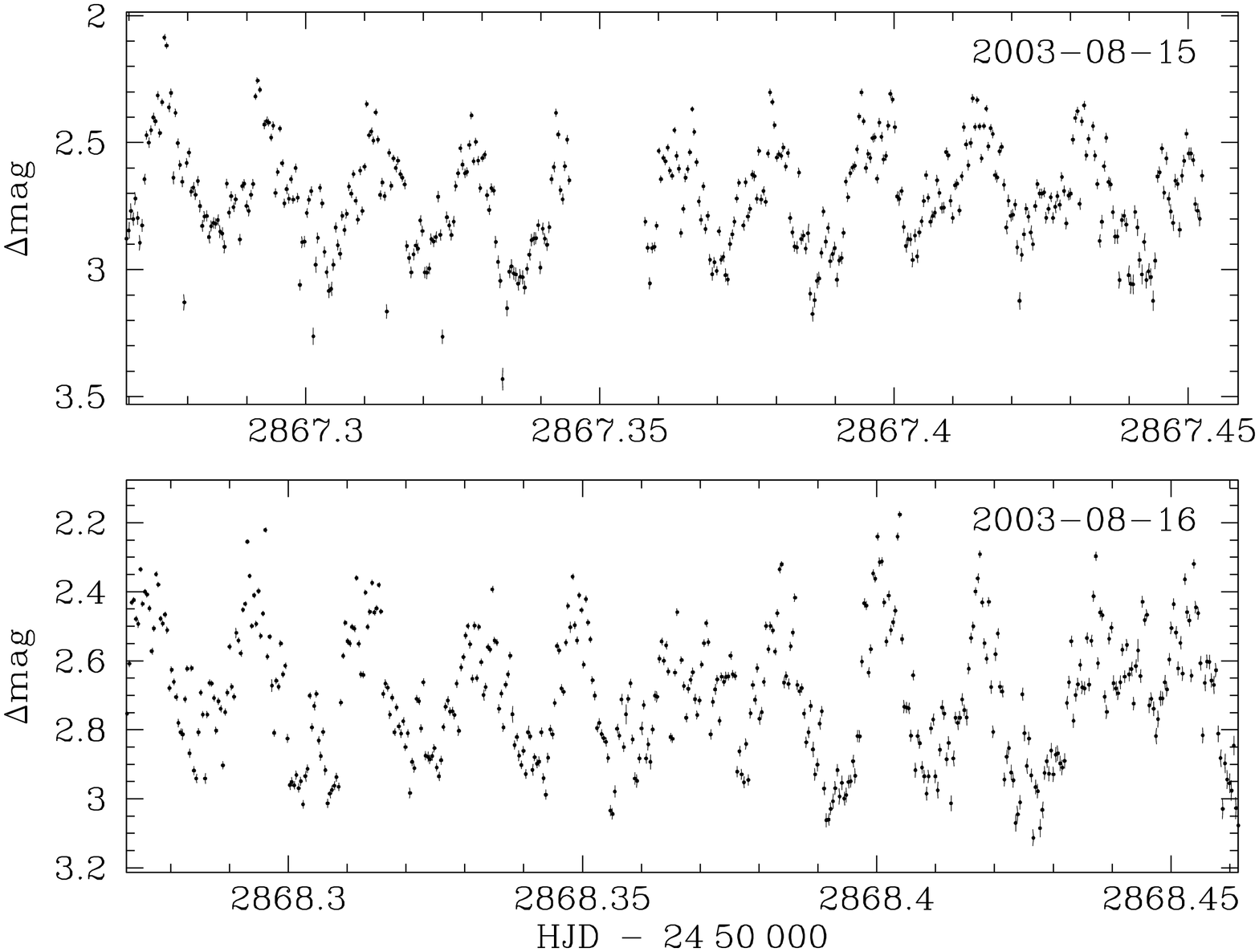}
\caption{\label{f-lc1803}Sample light curves of RX\,J1803 obtained at Kryoneri.}
\end{figure}

\begin{figure}
\includegraphics[angle=-90,width=\columnwidth]{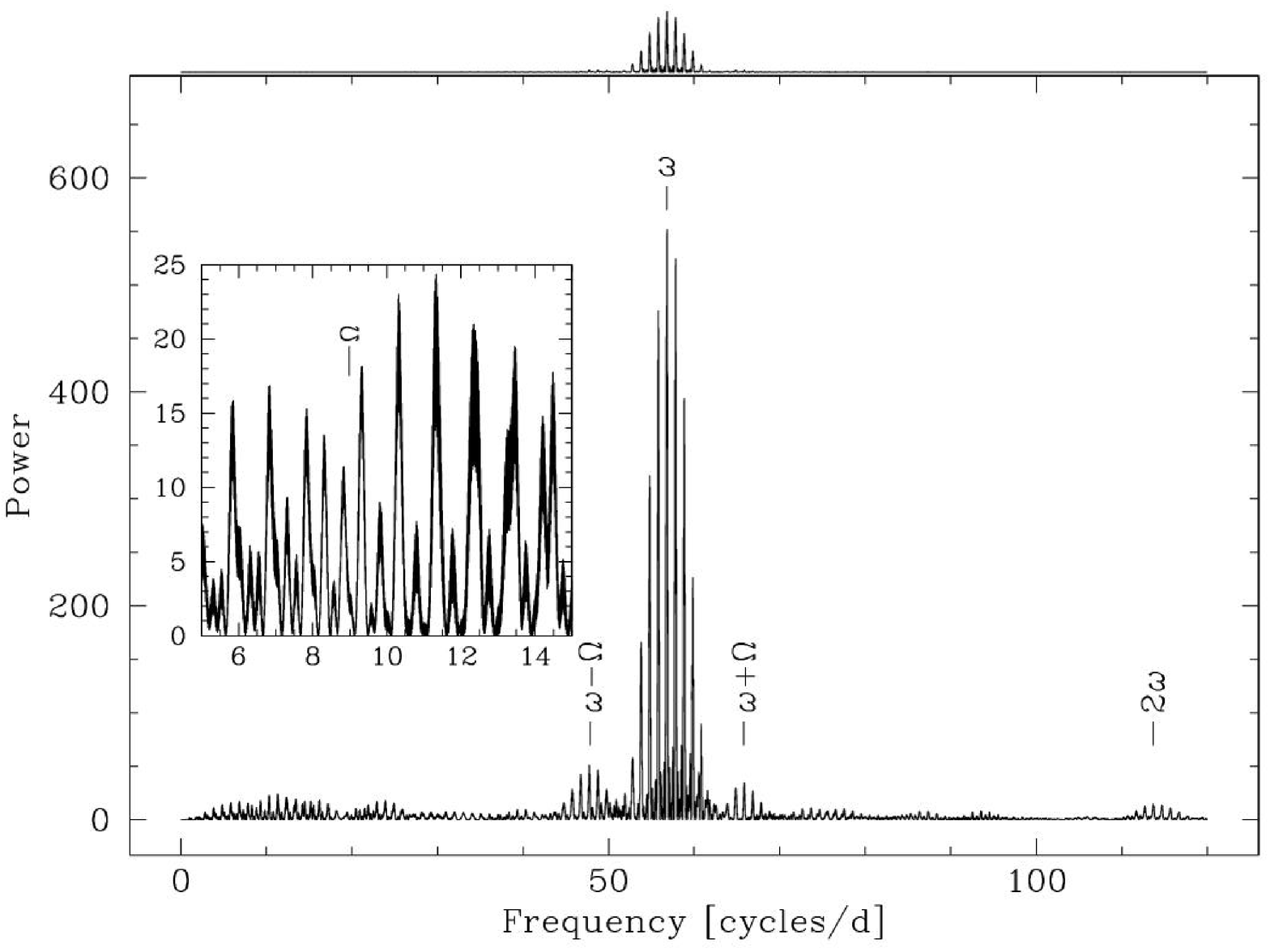}
\caption{\label{f-scargle1803}Scargle periodogram computed from the
  combined photometric data of RX\,J1803 (Table\,\ref{t-obslog}). Main
  panel: full frequency range. The orbital period and spin period are
  $\Porb=160.21$\,min and $\Pspin=1520.510$\,s. Small panel:
  enlargement around the orbital frequency. Shown on top is the window
  function calculated from the temporal sampling of the data, shifted
  to $\omega$.}
\end{figure}

\begin{figure}
\includegraphics[angle=-90,width=\columnwidth]{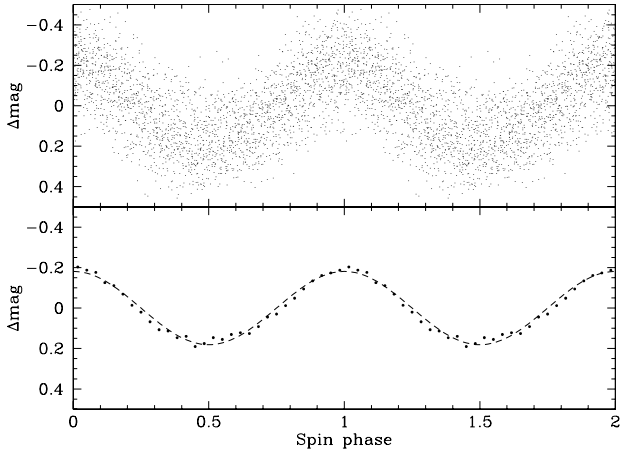}
\caption{\label{f-spinfold1803}Spin-folded light curves of RX\,J1803,
  using $\Pspin=1520.510$\,s. Top panel: all individual data points
  (Table\,\ref{t-obslog}). Bottom panel: Data binned into 30 phase
  slots. Plotted as a dashed line is a sine fit to the binned \&
  folded data.}
\end{figure}

\begin{figure}
\includegraphics[angle=-90,width=\columnwidth]{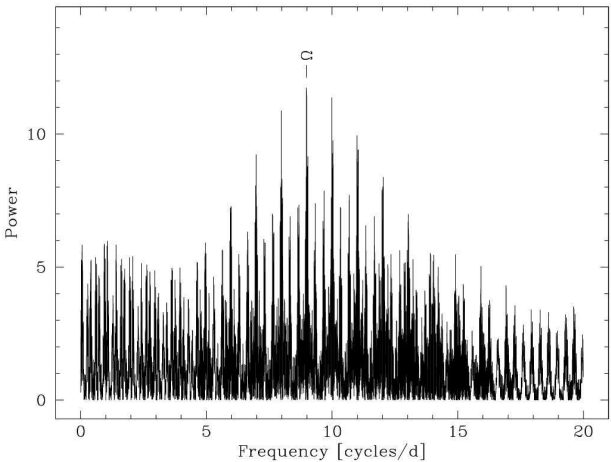}
\caption{\label{f-rvscargle1803} Scargle periodogram computed from the
  radial velocity variation of the H$\alpha$ line wings in RX\,J1803,
  determining the orbital period to be $\Porb=160.21$\,min. }
\end{figure}

\begin{figure}
\includegraphics[angle=-90,width=\columnwidth]{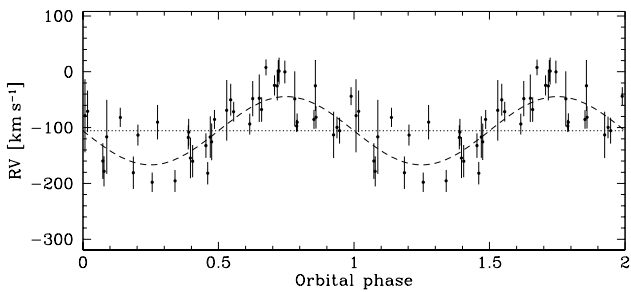}
\caption{\label{f-rvfold1803} Radial velocities (RV) determined from the
  H$\alpha$ wings in RX\,J1803, folded over the putative orbital
  period of 160.21(12)\,min. A sine fit to the folded radial
  velocities is shown as a dashed line. The systemic velocity determined from the
  fit is $\gamma=-100(7)$\,\kms.}
\end{figure}

\subsubsection{Photometry}
The light curves of RX\,J1803 unmistakably reveal the presence of
large-amplitude pulsations with a period of $\sim25$\,min
(Fig.\,\ref{f-lc1803}). The Scargle periodogram computed from the
photometric data confirms the coherent nature of these pulsations
(Fig.\,\ref{f-scargle1803}). A sine fit to the 5 consecutive nights of
Kryoneri August 2003 data results in $P=1520.510(66)$\,s. The first
harmonic of this signal is detected in the periodogram. Including the
single night of data obtained with the JKT in May 2003 in the time
series analysis results in substantial cycle count ambiguities due to
the long separation from the August data, and does not improve the
period determination.  Power in excess of the alias structure is
detected near 47.8\,\id, and to a lesser degree near 65.9\,\id. We
interpret the strongest signal as the signature of the white dwarf
rotation $\omega$, and the two weaker symmetrically offset signals as
orbital side bands, $\omega-\Omega$ and $\omega+\Omega$. Under this
assumption, the orbital period of RX\,J1803 is $\simeq159$\,min,
i.e. in the period gap. We note that no significant photometric
variability at the orbital period is detected in the periodogram.  The
times of spin maxima for each night of photometry, determined from
fitting a sine-wave to the individual data sets, are listed in
Table\,\ref{t-hjdzero} (where the period was fixed to the value
determined from the entire data set).

\subsubsection{Time-resolved spectroscopy}
The time-resolved spectroscopy obtained at Calar Alto and at the INT
(Table\,\ref{t-obslog}) provides the possibility for an independent
orbital period determination. For this purpose, we have rebinned all
spectra to a uniform wavelength scale, corrected for the heliocentric
velocity of the Earth, and convolved the continuum-normalised profiles
of H$\alpha$ and \Line{He}{II}{4686} with a double-Gaussian of fixed
width ($\mathrm{FWHM}=100\,\kms$) and separation 1400\,\kms. A Scargle
periodogram computed from the radial velocity measurements obtained in
this way displays the strongest peak at $f\simeq9.0$\,\id
(Fig.\,\ref{f-rvscargle1803}), flanked by 1-day aliases, which we
interpret as the signature of the orbital frequency. A sine fit to the
radial velocities results in $\Porb=160.21(12)$\,min, which coincides
with the orbital period estimated from the beat signals detected in
the photometry. The H$\alpha$ radial velocity data of RX\,J1803 folded
over the orbital period is shown in Fig.\,\ref{f-rvfold1803}.

In summary, RX\,J1803 is a bona-fide IP within the period gap,
$\Porb=160.21$\,min, and a likely white dwarf spin period of
$\Pspin=1520.510$\,s.

\section{Discussion}
We initiated a search for new CVs based on the X-ray/IR selection
described in Sect.\,\ref{s-selection} with the aim to identify
low-mass-transfer CVs close to the period minimum, such as WZ\,Sge,
BW\,Scl, or GD\,552, some of which might be period-bouncers. Our two
identification runs led to the discovery of 12 new CVs, representing
an impressive ``hit-rate'' of 7\,\%, but \textit{none} of the newly
identified systems resembles the short-period template
systems. Whereas these three template systems have RASS count rates
and $B$-band magnitudes comparable to the newly identified CVs
(WZ\,Sge: 0.2\,\cts, $B=15.2$; BW\,Scl: 0.18\,\cts, $B=15.6$, GD\,552:
0.18\,\cts, $B=16.6$), it may be that these three systems are
\textit{not} typical representatives of short-period low-mass-transfer
CVs, and may exhibit a higher X-ray emission. Moreover, the optical
flux of low-mass-transfer CVs is dominated by the emission of the
white dwarf. For an assumed white dwarf temperature of 15\,000\,K, our
brightness cut-off of $R=17$ corresponds to a maximum distance of
$\simeq125$\,pc. While this white dwarf temperature is appropriate for
WZ\,Sge, it may well be that the bulk of the short-period CVs have
cooler white dwarfs, for a white dwarf temperature of 10\,000\,K
$R=17$ corresponds to a maximum distance of $\sim85$\,pc only. The
conclusion from our survey at its current state is that there is not a
large population of quiescent nearby WZ\,Sge-like stars, and that a
large population of low-mass-transfer CVs can only exist if they look
different from WZ\,Sge and look-alikes. One example of a ``true''
low-mass-transfer CV is the recently identified HS\,2331+3905, which
contains indeed a cool ($\simeq10\,000$\,K) white dwarf and is not
detected in the RASS, although it is at a nearby distance of
$\simeq90$\,pc \citep{araujo-betancoretal05-1}.

A large fraction (two thirds) of the new CVs displays moderate to
strong \Line{He}{II}{4686} emission, suggestive of a magnetic
nature~--~not too surprising, as $\sim60\%$ of all X-ray discovered
CVs are either polars or IPs \citep{gaensicke05-1}. What is
surprising, however, is the very large fraction (one third) of new IPs
identified in this sample.

\begin{figure}
\includegraphics[width=\columnwidth]{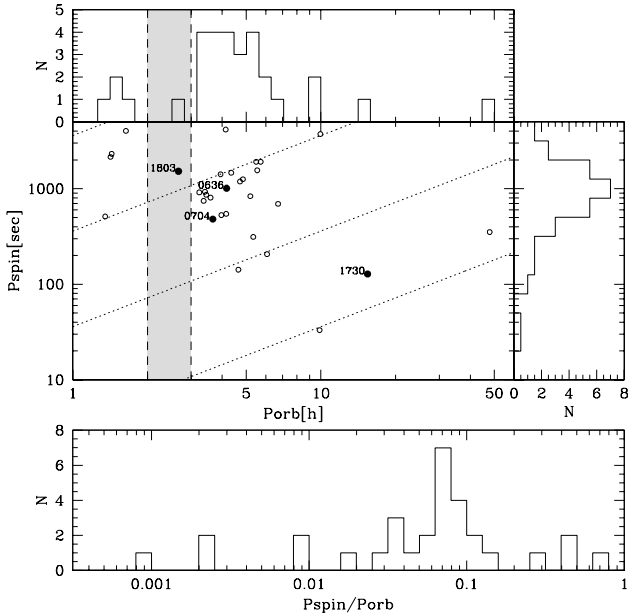}
\caption{\label{f-porb_pspin} Centre panel: Orbital and spin periods
of 31 confirmed IPs. The dotted lines indicate (from
top to bottom) $\Pspin/\Porb=1, 0.1, 0.01, 0.001$ . Top panel: Orbital
period distribution of the known IPs, the $2-3$\,h period gap is
shaded gray. Right panel: Spin period distribution of the known
IPs. The four new systems presented in this paper are shown as filled symbols.}
\end{figure}

\subsection{Intermediate polars: A growing family}
The census of IPs is probably the worst defined of all CV
subclasses. The IP page by K.
Mukai\footnote{http://lheawww.gsfc.nasa.gov/users/mukai/iphome/iphome.html}
lists 25 confirmed IPs. The latest edition (7.4) of
\citet{ritter+kolb03-1} has 37 IPs, and \citet{downesetal01-1}
contains 46 IPs. The defining criterion of an IP is the detection of
\textit{coherent} short-period variability over a sufficiently long
period of time, that can be associated with the white dwarf
spin~--~and it is the differing degrees of strictness in applying this
criterion which causes the large disagreement in the head count of
IPs. We adopted Mukai's conservative classification, and updating his
list by RX\,J0153.3+7446 ($\Pspin=1414$\,s,
\citealt{haberl+motch95-1}; $\Porb=236.48$\,min, Thorstensen,
priv.comm.) and HS0943+1404 ($\Pspin=4150$\,s, $\Porb\simeq250$\,min,
\citealt{rodriguez-giletal05-2}), as well as the four new IPs
presented here, the current roster of IPs encompasses 31
systems\footnote{Thus, our definition of IPs excludes systems showing
coherent variability at extremely short periods that lack the definite
evidence of these periods being related to the white dwarf spin
(e.g. WZ\,Sge; \citealt{robinsonetal78-1} and HS2331+3905;
\citealt{araujo-betancoretal05-1}), nearly-synchronous polars, systems
such as V795\,Her or LS\,Peg, i.e. members of the SW\,Sex class which
might contain a magnetic white dwarf \citep{rodriguez-giletal01-1} but
which lack undisputable measurements of $\Pspin$, and systems for
which the presence of coherent variability is supported only by
moderately small amounts of data.}.

Figure\,\ref{f-porb_pspin} shows the distribution of these 31 systems
in the $\Porb-\Pspin$ plane, their orbital and spin period
distributions, as well as the distribution in \Pspin/\Porb. A large
fraction of all IPs is found near $\Pspin/\Porb\simeq0.1$, which was
first noticed by \citet{barrettetal88-1}, and stimulated theoretical
work on the white dwarf spin equilibrium in magnetic CVs
\citep{king+lasota91-1, warner+wickramasinghe91-1}. Later,
\citet{king+wynn99-1} showed that spin equilibria with
$\Pspin/\Porb>0.1$ exist, explaining systems such as EX\,Hya. Based on
their calculations, \citet{king+wynn99-1} predicted that most IPs with
$\Pspin/\Porb>0.1$ should have orbital periods below the period
gap~--~which agrees with Fig.\,\ref{f-porb_pspin}.  Most recently,
\citet{nortonetal04-1} found that a large range of spin equilibria
exists in the ($\Pspin/\Porb$, $\mu_\mathrm{wd}$, \Porb) parameter
space, where $\mu_\mathrm{wd}$ is the magnetic moment of the white
dwarf. Assuming that the majority of the observed IPs are close to
their spin equilibria, \citet{nortonetal04-1} estimated their
$\mu_\mathrm{wd}$ from the model calculations. Using their Fig.\,2, we
estimate the magnetic moments of the white dwarfs in RX\,J0636,
RX\,J0704, and RX\,J1803 to be $\sim3\times10^{33}\mathrm{G\,cm^3}$,
$\sim1\times10^{33}\mathrm{G\,cm^3}$, and
$\sim4\times10^{33}\mathrm{G\,cm^3}$, respectively. In that respect,
RX\,J0636 and RX\,J1803 are similar to the bulk of the previously
known IPs, whereas RX\,J0704 resembles V405\,Aur or YY\,Dra.

The case of RX\,J1730 is quite different, its very low $\Pspin/\Porb$
and long orbital period suggest that the white dwarf is rotating much
faster than its equilibrium spin period, and we suggest that it is
similar to AE\,Aqr. In AE\,Aqr, the rapid white dwarf spin is actually
preventing accretion onto the white dwarf \citep{wynnetal97-1}, and
the mass-ejection from the system explains the large observed
spin-down rate of the white dwarf
\citep{dejageretal94-1}. \citet{schenkeretal02-1} suggested that
AE\,Aqr descended from a supersoft X-ray binary which only recently
($\sim10^7$\,yr ago) turned from a phase of extremely high
(thermal-timescale) mass transfer into a (not quite) ``ordinary''
CV. AE\,Aqr differs from normal CVs in having a spectral type which is
too late for its orbital period, and an extreme N/C
overabundance. \citet{gaensickeetal03-1} have shown that a substantial
fraction of CVs ($\sim10-15\%$) show evidence for enhanced N/C
abundances, confirming that thermal-time scale mass transfer may
represent an important CV formation channel. If RX\,J1730 is indeed a
sibling of AE\,Aqr, we predict that it should show a measurable
spin-down of the white dwarf rotation, and is likely to exhibit an
enhanced N/C abundance ratio as well. It may also show the flaring
characteristic of AE\,Aqr.

The class of IPs has seen a substantial growth in number over the last few
years, and more than 20\,\% of all confirmed IPs were identified by
us, DW\,Cnc (\,=\,HS0756+1624, literally a twin of V1025\,Cen,
\citealt{rodriguez-giletal04-1}, see also
\citealt{pattersonetal04-1}), 1RXS\,J062518.2+733433
(\,=\,HS0618+7336, a ``standard'' IP, 
\citealt{araujo-betancoretal03-2}), and HS0943+1404 (an IP with long
\Porb\ and small \Pspin/\Porb, \citealt{rodriguez-giletal05-2}) as part
of our search for CVs in the Hamburg Quasar Survey, and RX\,J0636,
RX\,J0704, RX\,J1730 and RX\,J1803 in the ROSAT/2MASS selected CV
sample described in this paper. These seven new IPs span a large range
in orbital and spin periods, with $0.0023<\Pspin/\Porb<0.43$,
demonstrating rich variety of systems among the newly identified CVs.

\subsection{INTEGRAL~--~a bloodhound for hunting down IPs?}
RXJ\,1730 has been detected as a significant INTEGRAL source
(IGR\,J17303--0601, $0.28\pm0.03$\,\cts) in the IBIS/ISGRI soft gamma
ray galactic plane survey (GPS)
\citep{birdetal04-1}. \citet{masettietal04-1} observed three optically
unidentified INTEGRAL sources, including RX\,J1730. Based on their
single spectrum of RX\,J1730, which is similar to our data shown in
Fig.\,\ref{f-idspec}, the authors suggested that the object is a
low-mass X-ray binary in the Galactic bulge. However, the detection of
a 128\,s spin period in our optical photometry clearly identifies
RX\,J1730 as an IP. Inspecting the IBIS GPS catalogue
\citep{birdetal04-1}, 6 out of the 123 sources are CVs, of which four
are IPs (1RXS\,J154814.5--452845, V2400\,Oph, RX\,J1730, and
V1223\,Sgr), one is an asynchronous polar (V1432\,Aql) and one is a
(non-magnetic) dwarf nova (SS\,Cyg). Whereas the dominant population
of ROSAT-discovered CVs were X-ray soft polars, the harder X-ray
spectrum of IPs makes them the most prolific CV type contained in
X-ray surveys of the Galactic plane, where interstellar absorption
will suppress the X-ray soft polars even for relatively short
distances. It appears that optical follow-up of unidentified INTEGRAL
GPS sources may turn out to be a significant discovery chanel for
IPs.

\begin{table}
\caption{\label{t-periods} Orbital and spin periods of the four new
  IPs.}
\begin{tabular}{lccrrrrrrcccccc}
\hline
System    & \Porb\,[min] & \Pspin\,[s] \\
\hline
RX\,J0636 & $\simeq201$  & 1008.3408(19) or 930.5829(40) \\
RX\,J0704 & $\simeq250$  & 480.7080(44)   \\
RX\,J1730 & 925.27(14)   & 127.999909(49) \\
RX\,J1803 & 160.21(12)       & 1520.510(66) \\
\hline
\end{tabular}
\end{table}

\section{Conclusions}
We have initiated a search for CVs selecting targets from the ROSAT
Bright Source catalogue and discriminating against AGN by using a
2MASS infrared colour-cut. Whereas our aim was to find members of the
``missing'' population of low-mass-transfer CVs near the minimum
period, we predominantly discovered new magnetic CVs, among which are four
IPs: RX\,J0636, RX\,J0704, RX\,J1730, and RX\,J1803. RX\,J0636
($\Porb\simeq201$\,min, $\Pspin=1008.3408$\,s or 930.5829\,s) 
and RX\,J1803 ($\Porb=160.21$\,min, $\Pspin=1520.510$\,s) are rather
typical IPs, except for RX\,J1803 being the first confirmed IP in the
period gap. RX\,J0704 ($\Porb\simeq250$\,min, $\Pspin=480.708$\,s)
appears to be similar to V405\,Aur or YY\,Dra. The most striking
new CV is RX\,J1730 ($\Porb=925.27$\,min, $\Pspin=128.0$\,s), which
may be a sibling of AE\,Aqr. RX\,J1730 is also a moderately bright
hard X-ray source in the INTEGRAL/IBIS Galactic plane survey. 

\section*{Acknowledgments}
BTG, TRM, and PRG were supported by a PPARC Advanced Fellowship, a
PPARC Senior Fellowship, and a PPARC PDRA, respectively. AE thanks the
Royal Society for generous funding. DS acknowledges a Smithsonian
Astrophysical Observatory Clay Fellowship.
Based in part 
on observations obtained at the German-Spanish Astronomical Center,
Calar Alto, operated by the Max-Planck-Institut f\"{u}r Astronomie,
Heidelberg, jointly with the Spanish National Commission for
Astronomy;
on observations made with the OGS telescope, operated on the island of
Tenerife by the European Space Agency (ESA) in the Spanish
Observatorio del Teide of the Instituto de Astrofisica de Canarias (IAC);
on observations made at the 1.2\,m telescope, located at Kryoneri
Korinthias, and owned by the National Observatory of Athens, Greece;
on observations made with the William Herschel Telescope, the Isaac
Newton Telescope, and the Jacobus Kapteyn Telescope, which are
operated on the island of La Palma by the Isaac Newton Group in the
Spanish Observatorio del Roque de los Muchachos of the IAC;
and on observations made with the 6.5m Magellan-Clay telescope 
operated at Las Campanas Observatory, Chile on behalf of the Magellan 
consortium; 
and on observations made with the Telescopio Nazionale Galileo 
operated on the island of La Palma by the Centro Galileo Galilei of
the INAF (Istituto Nazionale di Astrofisica) at the Spanish
Observatorio del Roque de los Muchachos of the IAC;
This publication makes use of data products from the Two Micron All
Sky Survey, which is a joint project of the University of
Massachusetts and the Infrared Processing and Analysis
Center/California Institute of Technology, funded by the National
Aeronautics and Space Administration and the National Science
Foundation.
We thank the referee Coel Hellier for his prompt report.
On 13 February 2005, during the write-up of this paper, our colleague
and friend Emilios Harlaftis died in an avalanche on Mt. Menalon,
Peloponnese. Emilios has been a very active astronomer, working mainly
on X-ray binaries and cataclysmic variables, and he strongly promoted
astronomy in Greece. He has involved many students actively in this
research projects through the use of Kryoneri observatory. We have
been working with Emilios for many years, and shared many moments with
him. We shall miss him.

\bibliographystyle{aa}
\bibliography{aamnem99,aabib}

\bsp

\label{lastpage}

\end{document}